\documentclass[letterpaper]{article} %
\usepackage[]{aaai25}  %
\usepackage{times}  %
\usepackage{helvet}  %
\usepackage{courier}  %
\usepackage[hyphens]{url}  %
\usepackage{graphicx} %
\usepackage{natbib}  %
\usepackage{caption} %
\usepackage{algorithm}
\usepackage{algorithmic}

\usepackage{newfloat}
\usepackage{listings}
\DeclareCaptionStyle{ruled}{labelfont=normalfont,labelsep=colon,strut=off} %
\floatstyle{ruled}
\newfloat{listing}{tb}{lst}{}
\floatname{listing}{Listing}
\usepackage{newfloat}
\usepackage{listings}
\usepackage{subfigure}
\usepackage{hhline,booktabs,colortbl,multirow,tabularx,diagbox,tablefootnote,threeparttable} %
\usepackage{multirow}
\usepackage{prettyref}
\usepackage{xcolor}
\usepackage[shortlabels]{enumitem}
\usepackage{float}
\usepackage{natbib}
\usepackage{amsthm,amsmath,amssymb,mathtools}
\definecolor{mycyan}{gray}{.7}

\newrefformat{fig}{Figure~\ref{#1}}
\newrefformat{tab}{Table~\ref{#1}}
\newrefformat{sec}{Section~\ref{#1}}
\newrefformat{app}{Appendix~\ref{#1}}
\newrefformat{alg}{Algorithm~\ref{#1}}
\newrefformat{property}{Property~\ref{#1}}
\newrefformat{theorem}{Theorem~\ref{#1}}
\newrefformat{corollary}{Corollary~\ref{#1}}
\newrefformat{proposition}{Proposition~\ref{#1}}
\newrefformat{def}{Definition~\ref{#1}}
\newrefformat{eq}{equation~(\ref{#1})}
\newrefformat{app}{Appendix~\ref{#1}}
\newcommand{\pref}{\prettyref}

\def\our{\texttt{OmniGenome}}
\def\ourplus{\texttt{OmniGenome$+$}}
\def\oursmall{\texttt{OmniGenome}$^{52\texttt{M}}$}
\def\ourbase{\texttt{OmniGenome}$^{186\texttt{M}}$}
\def\oursmallplus{\texttt{OmniGenome$^{52\texttt{M}}+$}}
\def\ourbaseplus{\texttt{OmniGenome$^{186\texttt{M}}+$}}

\def\strtoseq{Str2Seq}
\def\seqtostr{Seq2Str}

\title{Bridging Sequence-Structure Alignment in RNA Foundation Models}
\author{
	Heng Yang\textsuperscript{\rm 1}, Renzhi Chen\textsuperscript{\rm 2}, Ke Li\textsuperscript{\rm 1}\thanks{Corresponding Author}\\
}
\affiliations{
	\textsuperscript{\rm 1} Department of Computer Science,
	University of Exeter, EX4 4QF, Exeter, UK \\
	\textsuperscript{\rm 2} Qiyuan Lab, Beijing, China \\
	\{hy345, k.li\}@exeter.ac.uk, \{chengrenzhi1989\}@gmail.com
}

\usepackage{bibentry}

\begin{document}

\maketitle

\begin{abstract}
The alignment between RNA sequences and structures in foundation models (FMs) has yet to be thoroughly investigated. Existing FMs have struggled to establish sequence-structure alignment, hindering the free flow of genomic information between RNA sequences and structures. In this study, we introduce \our, an RNA FM trained to align RNA sequences with respect to secondary structures based on structure-contextualised modelling. The alignment enables free and bidirectional mappings between sequences and structures by utilising the flexible RNA modelling paradigm that supports versatile input and output modalities, i.e., sequence and/or structure as input/output. We implement RNA design and zero-shot secondary structure prediction as case studies to evaluate the \seqtostr\ and \strtoseq\ mapping capacity of \our. Results on the EternaV2 benchmark show that \our\ solved $74\%$ of puzzles, whereas existing FMs only solved up to $3\%$ of the puzzles due to the oversight of sequence-structure alignment. We leverage four comprehensive \textit{in-silico} genome modelling benchmarks to evaluate performance across a diverse set of genome downstream tasks, where the results show that \our\ achieves state-of-the-art performance on RNA and DNA benchmarks, even without any training on DNA genomes.
\end{abstract}

\section{Introduction}
\label{sec:intro}

RNA is a critical type of molecule that encodes a vast array of biological regulatory elements that orchestrate crucial aspects of plant growth, development, and adaptation to environmental stresses. To decipher the genomic code in RNA and manipulate RNA engineering and design, current research mainly uses bioinformatics in solving RNA genome-oriented challenges. Recent advancements in large-scale pre-trained foundation models (FMs) have demonstrated their unprecedented potential to back up existing genome analysis, as FMs are capable of learning and predicting the complex `genomic language'~\cite{Nguyen23} hidden in genome encoding processes. Existing FMs have been widely employed as basic sequence feature extractors to improve the performance of diverse genome analysis tasks, such as secondary structure prediction~\cite{TanFSM17,DanaeeRWDHH18,Mathews19,Kalvari21}, degradation rate prediction~\cite{Yaish22,Wayment22}, and mRNA vaccine design~\cite{Corbett20,Runge23}. In RNA, it is intriguing that the functionality and stability are intertwined with its complex structures in molecular biology~\cite{Ganser19}. However, the role of the structure as a second `genomic language' to interact with sequences and solve various RNA downstream tasks has been largely ignored.

\begin{figure}[htbp]
    \centering
    \includegraphics[width=\linewidth]{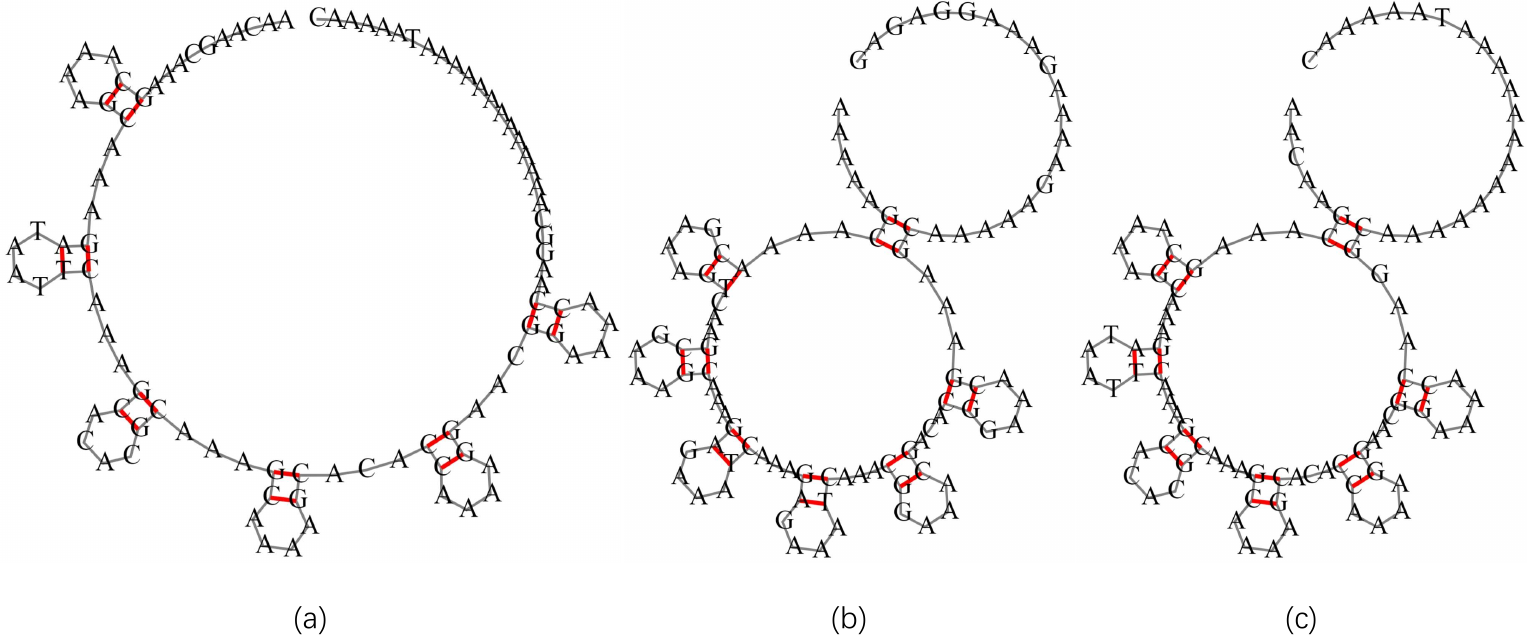}
    \caption{An example for \textit{in-silico} RNA folding drawn by ViennaRNA. The subfigures (a) and (c) indicate the same sequence with different structures. The subfigures (b) and (c) denote the identical structure can be from different sequences.}
    \label{fig:structure_example}
    \vspace{-15pt}
\end{figure}

\paragraph{Sequence-Structure Alignment in GFMs}

We define alignment between sequences and secondary structures\footnote{In this paper, all structures are referred to as the secondary structures.} as the bidirectional information flows. Current FMs have been struggling to establish an alignment between RNA nucleotide sequences and their folded structures, thus impeding bidirectional genomic information flows. 
There has been a deep scientific challenge to align RNA sequences with structures because it is not deterministic to predict sequences from structures and vice versa. In other words, an identical sequence may be folded into different sub-optimal structures because the folding patterns of RNA sequences depend on various \textit{in-vivo} factors~\cite{Tinoco99}. Further, a structure can be folded from different sequences composed of variational combinations of nucleotide bases, as the example shown in~\pref{fig:structure_example}. 
The oversight of such alignment in existing FMs causes outstanding issues in understanding and leveraging RNA structures, such as mRNA design. For example, recent state-of-the-art RNA FMs, RNA-FM ~\cite{Chen22} and RNA-MSM ~\cite{ZhangLJGXLCSHXS24}, only solved $3$ out of $100$ puzzles in \textit{in-silico} RNA design ~\cite{Lee14}. This is because they fail to decipher corresponding sequences based on structures to guide RNA design. 

To address the above two problems, we propose sequence-structure alignment in RNA FMs, which leverages the large-scale annotations of sequences and structures to build reliable structure to sequence (\strtoseq) and sequence to structure (\seqtostr) mappings, leading to an \textit{aligned} FM dubbed \our. The sequence-structure alignment enables genomic information to freely flow between sequences and structures by introducing a flexible RNA modelling paradigm that supports versatile inputs and outputs modalities, i.e., sequence and/or structure as input/output.
The sequence-structure alignment enables genomics information to freely flow between sequences and structures by introducing a flexible RNA modelling paradigm that supports versatile inputs and outputs modalities, i.e., sequence and/or structure as input/output. Furthermore, the sequence-structure alignment is designed to be architecture-agnostic and genome-agnostic. That is to say, it can be easily transferred to large-scale models with new architecture and different genome types like DNA. 

\begin{figure}[htbp]
    \centering
    \includegraphics[width=\linewidth]{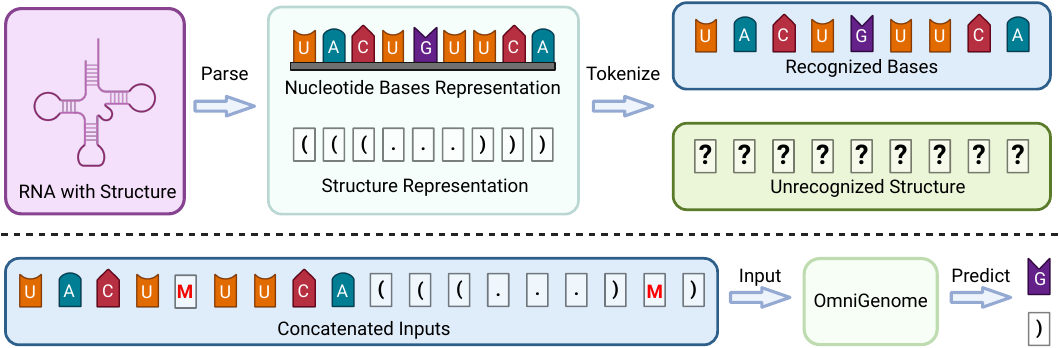}
    \caption{A virtual example of structure-contextualised sequence reconstruction. The top subfigure indicates that we need to expand the vocabulary for structure-aware tokenization. Otherwise, the structure cannot be recognised, i.e., unknown as ``\textbf{?}''. We show our structure-contextualised modelling (\strtoseq) in the bottom sub-figure, where the `\textcolor{red}{M}' indicates the masked tokens to be reconstructed by \our.}
    \label{fig:alignment}
    \vspace{-10pt}
\end{figure}

\paragraph{\strtoseq\ Mapping}

RNA structure serves as a vital input in most of the RNA genome analysis tasks.
To induce the ability of \strtoseq\ mapping in genomic FMs, we formulate a structure-contextualised RNA sequence reconstruction task, which stems from the representation of RNA secondary structures in texts composed of dots and brackets. As the diagram is shown in~\pref{fig:alignment}, we first concatenate sequence-structure pairs as inputs and then mask a small portion of nucleotide bases in the sequence. Then, we pre-train \our\ to reconstruct the masked nucleotide bases given the structure contexts. This simple but effective formulation of \strtoseq\ mapping realises structure input awareness in genomics FM pre-training and provides substantial compatibility for structure-contextualised tasks, which has been verified in our RNA design benchmark.

\paragraph{\seqtostr\ Mapping} On the other hand, \seqtostr\ mapping, such as end-to-end secondary structure prediction (SSP)~\citep{Sato21MXFold2,Fu22UFold}, is another critical aspect of achieving the alignment. We generalise end-to-end structure pre-training~\citep{Yan22RSSMFold} to \our\ pre-training. This large-scale structure pre-training on diversified genomes supervises \our\ to perform \seqtostr\ mapping. The problem of structure pre-training lies in RNA structure annotation scarcity, which leads to biased structure predictions~\citep{Chen20E2EFold} and barriers the structure prediction robustness on small datasets. 
To conduct \seqtostr\ mapping, tremendous secondary structure annotations are required to avoid data bias. A feasible solution to RNA structure pre-training is leveraging the plausible structures calculated based on the minimum free energy. In this paper, we leverage the popular ViennaRNA~\cite{Lorenz11} to serve our purpose, \lq computing\rq\ the structures for millions of RNA sequences and perform structure pre-training in \our. 

\paragraph{Evaluations and Results} To validate the effectiveness of \our, we designed four large-scale genome benchmarks with diverse genomics tasks. The first one is the RNA genomics benchmark (RGB) compiled in the study, which contains diverse challenging genomics understanding tasks that benefit from the sequence-structure alignment, such as degradation rate prediction. The second benchmark is the plant genomics benchmark (PGB)~\cite{Mendoza23} which contains millions of DNA sequences to evaluate the DNA sequence understanding tasks. In particular, we want to use this benchmark to evaluate the generalisability of \our\ among diversified species and genomes. The overall performance of \our\ (up to $186$M parameters) on both two benchmarks consistently outperforms existing genomics FMs with up to $35\%$ improvement, even compared with Agro-NT~\cite{Mendoza23} which contains $1$ billion parameters. The last two benchmarks, available in the appendix, are the genomics benchmark (GB)~\cite{Grevsova23} and genomics understanding evaluation (GUE)~\cite{Zhou23}, which serve as two additional DNA benchmarks to evaluate generalisability on non-plant genome modelling.  

In addition, we also conduct zero-shot \seqtostr\ and \strtoseq\ prediction experiments to verify the performance of sequence-structure alignment. As revealed in the experiments in Sections~\ref{sec:str2seq_exp} and~\ref{sec:seq2str_exp}, \our\ achieves up to a $74.85\%$ macro-F1 score in zero-shot \seqtostr\ prediction, i.e., secondary structure prediction, outperforming fine-tuned FMs and bioinformatics methods like ViennaRNA. In terms of \strtoseq\ prediction performance, we evaluate the performance of \our\ in the \textit{in-silico} RNA design task. We solved $74\%$ of complex puzzles of the EternaV2 benchmark~\cite{Lee14}, while state-of-the-art FMs such as RNA-MSM and RNA-FM only solved up to $3\%$. Besides, \our\ only takes less than one hour to solve most of the puzzles, while most RNA design methods need to take up to $24$ hours to solve even a single puzzle. 

\paragraph{Open-source Toolkit and Tutorials} Open science is always the golden standard to promote this rising area of FM for genome modelling, which unfortunately lacks relevant high-quality resources such as code integrity, data availability, and pre-training pipeline. To address this gap, following the FAIR principles~\cite{Wilkinson16fair}, we developed an open-source package\footnote{\url{https://github.com/yangheng95/OmniGenBench}} that includes step-by-step tutorials for FM pre-training and downstream tasks fine-tuning, to name a few. It provides ready-to-use genomics benchmarks and uses the API with only a few lines of code to streamline benchmarking purposes. We believe this will be a valuable resource to make this emerging AI for the RNA community to thrive.

\section{Methodology}
\label{sec:method}

This section delineates the implementation details of \our\ including its entire pre-training workflow and downstream benchmarks.

\subsection{RNA Tokenization for Alignment}
\label{sec:tokenization}

\begin{figure}[htbp]
    \centering
    \includegraphics[width=\linewidth]{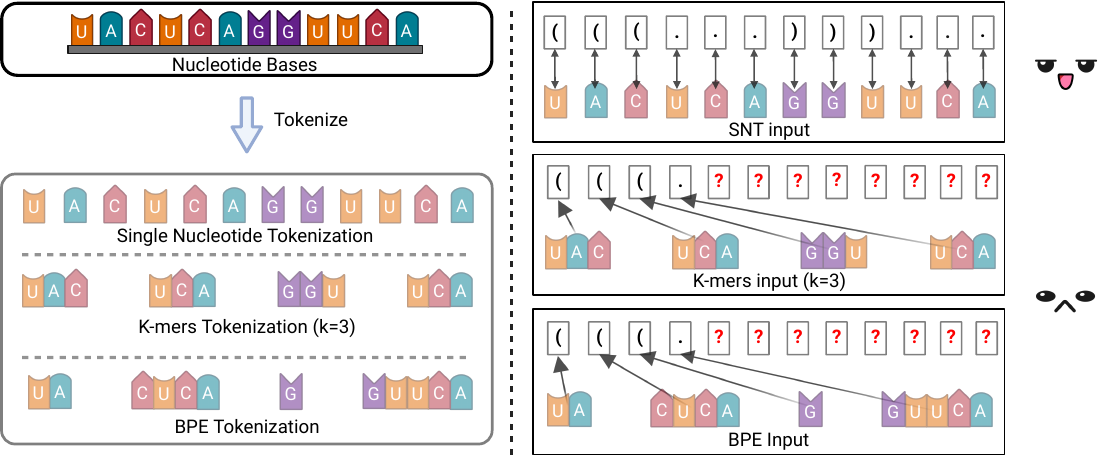}
    \caption{An illustrative example of RNA tokenization. The left sub-figure shows that k-mers and BPE entangle the bases and fail to align the SN-level inputs and outputs. The right sub-figure denotes that only SNT can achieve sequence-structure alignment, such as \seqtostr\ prediction.}
    \label{fig:tokenization}
    \vspace{-10pt}
\end{figure}

We aim to implement a fine-grained alignment between RNA sequences and structures, where each base in the sequences reflects a structural label in $\{$ `\texttt{(}', `\texttt{)}', `\texttt{.}'$\}$.
Therefore, we propose an adapted implementation of the single nucleotide tokenization (SNT) method ~\citep{Nguyen23,ChenZD23} in \our, where the whole vocabulary, $\{$\lq\texttt{A}\rq, \lq\texttt{T}\rq, \lq\texttt{C}\rq, \lq\texttt{G}\rq, \lq\texttt{U}\rq, \lq\texttt{N}\rq, \lq\texttt{(}\rq, \lq\texttt{)}\rq, \lq\texttt{.}\rq$\}$, contains the nucleotide-level structural labels. We illustrate our tokenization based on an example shown in \pref{fig:tokenization}.

Our adapted SNT features bidirectional mappings between single nucleotide (SN) bases and structural labels required by sequence-structure alignment. Another reason for the adaption of SNT is that,
in the realm of RNA genome modelling, the FM performance highly depends on the tokenization resolution ~\citep{Nguyen23,ChenZD23}. For example, the k-mers~\citep{YangLP23,Dalla23} and BPE~\citep{DevlinCLT19,Zhou23} tokenization methods combine multiple bases into tokens and embeddings, which compromise modelling resolution and thus fail to the solution of fine-grained genomic tasks like structure prediction as well as base-level degrade rate prediction. Like other encoder-only models, e..g, BERT~\citep{DevlinCLT19}, we incorporated special tokens, e.g., \lq\texttt{<mask>}\rq, to implement masked language modelling.

\begin{figure*}[t!]
    \centering
    \includegraphics[width=\linewidth]{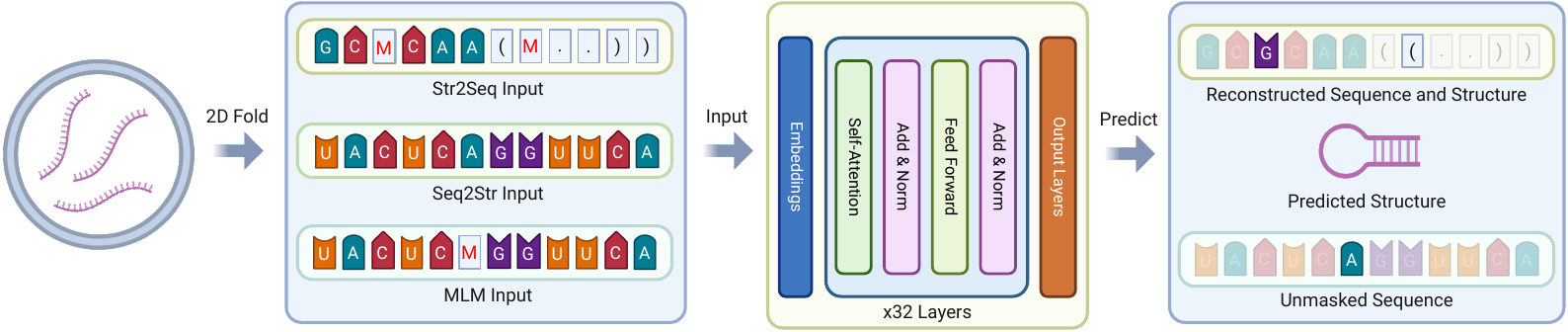}
    \caption{The workflow of \our\ pre-training. We craft the inputs for three pre-training objectives described in \pref{sec:objectives}. The outputs are reconstructed sequences based on the context of structure, predicted secondary structure, and unmasked sequences, respectively. The predictions of shadowed tokens are not calculated in the objective functions.}
    \label{fig:workflow}
    \vspace{-10pt}
\end{figure*}

\subsection{Pre-training Objectives}
\label{sec:objectives}

As discussed in ~\pref{sec:intro}, a key desideratum for SN-level genome modelling is to build the alignment between RNA sequences with corresponding secondary structures. Bearing this in mind, we formulate two pre-training objectives, i.e., $\mathcal{L}_\texttt{Str2Seq}$ and $\mathcal{L}_\texttt{Seq2Str}$, for \strtoseq\ and \seqtostr\ predictions, respectively. Besides, we aggregate these two objectives with the masked RNA language modelling objective \texttt{MRLM} to pre-train \our\ as follows: 
\begin{equation}
\mathcal{L}_\texttt{pre-train}=\mathcal{L}_\texttt{Str2Seq}+\mathcal{L}_\texttt{Seq2Str}+\mathcal{L}_\texttt{MRLM}+\lambda||\theta||_2,
    \label{eq:pretraining_objective}
\end{equation}
where $\lambda$ is the $\ell_2$ regularisation weight and $\theta$ represents the parameters of \our. The following paragraphs explain the design principles of each objective function used in~\pref{eq:pretraining_objective}.

\begin{itemize}
    \item $\mathcal{L}_\texttt{Str2Seq}$ is designed to enable \our\ to predict bases given structure-contextualised sequences with partially masked bases. This objective aims at \strtoseq\ tasks and teaches \our\ to interpret structure information and infer the masked sequences. To achieve this objective, we mask $15\%$ of the bases and structure tokens, encouraging the model to infer masked bases (i.e., $\{$\lq\texttt{A}\rq, \lq\texttt{T}\rq, \lq\texttt{C}\rq, \lq\texttt{G}\rq, \lq\texttt{U}\rq, \lq\texttt{N}\rq$\}$) and structure tokens (i.e., $\{$\lq\texttt{(}\rq, \lq\texttt{)}\rq, \lq\texttt{.}\rq$\}$). Specifically, $\mathcal{L}_\texttt{Str2Seq}$ is defined as the classic cross-entropy loss widely used in the masked language modelling:
    \begin{equation}
        \mathcal{L}_\texttt{Str2Seq}=-\frac{1}{|m|}\sum_{i=1}^m\log p(x_i\mid x_{\setminus i}),
    \end{equation}
    where $m$ is the number of masked nucleotide and structure tokens, and $p(x_i|x_{\setminus i})$ indicates the probability of predicting the masked nucleotide $x_i$ based on its context.
    \item In terms of structure-out modelling, we implement $\mathcal{L}_\texttt{Seq2Str}$ to enable \our\ for \seqtostr\ predictions. Instead of directly feeding the secondary structure into \our\ as inputs, this objective employs the RNA secondary structures as labels for supervised training. This objective is implemented as a token-level classification, where the $\mathcal{L}_\texttt{Seq2Str}$ loss is defined in the following cross-entropy loss:
    \begin{equation}
        \mathcal{L}_\texttt{Seq2Str}=-\sum_{i=1}^N\sum_{c=1}^C s_{ic}\log(\hat{s}_{ic}),
    \end{equation}
    where $s_{ic}$ denotes the label $c$ of secondary structure at the $i$-th position, and $\hat{s}_{ic}$ is the probability predicted by a linear classifier deployed on \our. $N$ is the length of an RNA sequence and $C=3$ denotes the number of the possible labels of structure, i.e., $\{$\lq\texttt{(}\rq, \lq\texttt{)}\rq, \lq\texttt{.}\rq$\}$.
    \item The last objective,  $\mathcal{L}_\texttt{MRLM}$, is adapted to the conventional masked language modelling loss in NLP. It aims to improve the model's understanding of genomic language in RNA sequences by predicting the masked or replaced $5\%$ of nucleotide bases. The definition of $\mathcal{L}_\texttt{MRLM}$ is similar to that of $\mathcal{L}_\texttt{Str2Seq}$ which only considers the prediction of masked bases rather than randomly replaced bases. The loss function of MRLM is well-known so we omit its formula here.
\end{itemize}

We cannot trust structure predictions (in $\mathcal{L}_\texttt{Seq2Str}$) while the structures are leaked in inputs (in $\mathcal{L}_\texttt{Str2Seq}$), i.e., the sequence inputs and outputs of these two objectives are exclusive. In practice, we only consider objectives either $\mathcal{L}_\texttt{Seq2Str} + \mathcal{L}_\texttt{MRLM}$ or $\mathcal{L}_\texttt{Str2Seq} + \mathcal{L}_\texttt{MRLM}$ for each input sequence. In the pre-training, $70\%$ of RNA sequences are used for the first two objectives, while the remaining $30\%$ are used for the latter two objectives. This proportion setting is concluded from our empirical experience to balance the capability of \strtoseq\ and \seqtostr\ predictions.

\subsection{Model Architecture}
\label{sec:architecture}
 
\our\ adopts the Transformer encoder architecture with bidirectional multi-head attention. We do not adopt recent architectures like Mamba ~\cite{Gu23mamba,Schiff24caduceus} and Hyena~\cite{Nguyen23} because our experiments in \pref{tab:pgb_results} and \pref{tab:rgb_results} show that these architectures are not competent at RNA genome understanding. This low performance is probably because RNA sequences are much shorter than DNA sequences in the wild.

We designed two variants, dubbed \oursmall\ and \ourbase\ with $52$ and $186$ million parameters respectively. Some key model specifications are summarised in~\pref{tab:specification}.

\begin{table}[htbp]
\centering
\caption{Summary of some key model specifications of two \our\ variants. ``${*}$'' means that we used a modelling length of $1024$ in the pre-training, while the supports up to $4096$ in downstream tasks. }
\resizebox{.55\linewidth}{!}{
\begin{tabular}{c|cc}
\toprule
\our & \texttt{52M} & \texttt{186M} \\
\midrule
\# of Layers & $16$ & $32$ \\
Embedding dimension & $480$ & $720$ \\
Intermediate dimension & $2,400$ & $2,560$ \\
\# of heads & $24$ & $30$ \\
\# of parameters & $52$M & $186$M \\
Modelling length & \multicolumn{2}{c}{$4,096^{*}$} \\
\bottomrule
\end{tabular}
}
\label{tab:specification}
\vspace{-10pt}
\end{table}

To improve the reproducibility of \our, we list the pre-training settings and hyperparameters as follows. 
\begin{itemize}
    \item The learning rate is set to $5\times 10^{-5}$ and the weight decay is set to $0.01$. \item  We use \texttt{AdamW} as the optimiser with hyperparameters $\beta_1 = 0.9$ and $\beta_2=0.999$.
    \item  We use a linear decay strategy with a warm-up period of $1,000$ steps in the learning rate scheduler.
    \item The batch size is set to $2,048$.
    \item No dropout is applied during pre-training, and we use the rotary position embeddings ~\citep{SuALPBL24} to further enhance the model's scalability to long RNA sequences.
    \item We built a distributed training environment with $8$ \textsc{Nvidia} RTX $4090$ GPUs, while its configuration is introduced in ~\pref{app:environment}. The pre-training was finished in approximately 1 and 3 weeks for \oursmall\ and \ourbase, respectively.
\end{itemize}

\subsection{Pre-training Database: OneKP}
\label{sec:curation}
Recent studies ~\citep{ChenZD23,Zhou23} have shown that data diversity can enhance FM performance without significantly increasing model capacity. For the \our\ pre-training, we collected transcriptome data from the OneKP initiative\footnote{\url{https://sites.google.com/a/ualberta.ca/onekp/}}~\citep{Carpenter19}, which compiles large-scale RNA raw sequence database from $1,124$ plant species. The raw sequences are not available for pre-training before processing and filtering. 

We adopt the following raw sequence data curation protocol to fit pre-training. 

\begin{itemize}
    \item To enhance training efficiency and reduce bias, we removed all duplicate sequences. 
    \item To tackle incomplete transcriptome data and other noises, we discard sequences shorter than $50$ bases. 
    \item To facilitate the sequence-structure alignment training, we adopt ViennaRNA\footnote{\url{https://github.com/ViennaRNA/ViennaRNA}} to obtain the secondary structures for the sequences.
    \item We use cd-hit-est ~\cite{Li06cdhit} and blast ~\cite{Altschul90Blast} tools to filter the sequences in downstream tasks with similar structures. Please refer to the experiment section for more details.
\end{itemize}

\subsection{Benchmark Suites}
\label{sec:downstream_tasks}

\our\ is designed as a general-purpose RNA FM that can be fine-tuned for a diverse set of downstream genomics predictive tasks. In this paper, we constructed a large-scale benchmark suite for RNA FMs. According to the category of genomes, we split the benchmark into two parts.

\paragraph{RNA Genomic Benchmark (RGB)}
RGB is a collection of genome understanding tasks, as shown in \pref{tab:rgb_detail}. RGB contains $6$ SN-level tasks that are curated in this work or collected from published articles. The purpose of RGB is to benchmark FMs in challenging SN-level modelling tasks like the predictions of mRNA degradation rates and secondary structures. The sequence length in RGB ranges from $107$ to $512$, which is enough for most RNA understanding tasks. These multi-species and SN-level tasks in RGB serve as the first comprehensive RNA benchmark to assess the modelling capabilities of FMs. For detailed information on each dataset, such as their sources and sizes, please refer to \pref{app:rgb}.

\paragraph{Plant Genomic Benchmark (PGB)}
PGB\footnote{https://huggingface.co/datasets/InstaDeepAI/plant-genomic-benchmark}~\citep{Mendoza23} ) shown in \pref{tab:pgb_detail} provides a large-scale and comprehensive suite of DNA genome tasks designed to evaluate the modelling capabilities of FMs in plant biology. PGB involves $8$ types of DNA downstream subtasks, including a range of critical tasks such as promoter strength prediction and gene expression regression. There are $28$ datasets in total with millions of DNA sequences to be evaluated in PGB, and the sequence lengths are up to $6000$, which is very long for most of the genomic FMs. Since the original evaluation protocol is not publicly available, we have re-implemented the auto-benchmark for all the subtasks from PGB in our package. By integrating diverse downstream tasks, PGB aims to facilitate the development of plant genomics and robust assessment. Due to computational limitations, we randomly sample a maximum of $10$k examples in all downstream datasets in PGB to evaluate the FM's performance. 

\subsection{\strtoseq\ Modelling Case: RNA Design}
\label{sec:rna_design}
One of the most challenging practices addressed by \our\ is RNA design, which has not been settled in existing FMs because of the oversight of sequence-structure alignment. In this section, we provide a case of RNA design that requires the exploitation of sequence-structure alignment. To address RNA design, we introduce a simple but effective genetic algorithm (GA)\footnote{The experiment script is available at: \url{https://tinyurl.com/RNADesign}} to search feasible RNA candidates, which is based on the \strtoseq\ prediction capability of \our. The GA implementation details main steps in the genetic algorithm and workflow visualisation are available in \pref{app:rna_design} and \pref{fig:design}, respectively. In \pref{sec:seq2str_exp}, the experimental results on the Eterna V2 dataset indicate an impressive performance in RNA design compared to existing methods.

\section{Experiments}

\label{sec:experiments}
To evaluate the performance of \our\ across genome modelling, we implement experiments on diverse downstream tasks. We first evaluate the sequence-structure alignment capability of \our. Subsequently, we evaluate the overall performance of \our\ on two comprehensive genomic modelling benchmarks, i.e., RGB and the PGB, respectively. Finally, we include the GB and GUE in the appendix to evaluate the performance on non-plant genomes.

\subsection{RNA Sequence Filtering}
\label{sec:filtering}
The pertaining involves RNA sequences and structures prediction, we take the data and annotation leakage problem seriously.
\begin{itemize}
    \item To avoid structure annotation leakage of downstream benchmarks, the secondary structure predictors for all FMs were randomly initialised for fair comparisons, which means the pre-trained structure predictor of \our\ was not used in benchmarks, except for zero-shot SSP experiments. Please find the source codes for details.
    \item To reduce sequence leakage caused by evolutionary conservative sequences across multiple species, we use the ch-hit-est tool to calculate the sequence similarity between sequences from the OneKP database and downstream tasks. We adopt the similarity threshold of $80\%$ for ch-hit-est to eliminate sequences whose homogeneous sequences appeared in the OneKP database. Subsequently, we exploit the blastn tool to query potentially leaked sequences in downstream benchmark datasets and further alleviate the data leakage problem. The e-value has been set to $1$ for rigorous sequence filtering. 
\end{itemize}

\subsection{Comparison Baselines}
\label{sec:baselines}
Apart from \our, we implement a plus variant, i.e., \ourplus. In the context of \ourplus, we assume the structure annotation from ViennaRNA is always available for enhancing the model based on structure-contextualised modelling. In SSP tasks, we can also use the ViennaRNA's structure annotations as contexts to improve downstream SSP performance.
Please refer to \pref{app:baselines} for brief introductions of these FMs. 

We can compare \our\ with the following RNA and DNA FMs shown in \pref{tab:fm_details} as baselines to help evaluate the performance of \our. 
We are aware that some FMs are also developed for RNA, such as Uni-RNA~\citep{WangGCLJKW23}, 5UTR-LM~\citep{Chu24}, etc. However, we cannot compare \our\ with them because their source codes are very hard to work with in our efforts or not publicly available.

\subsection{Sequence-Structure Alignment Evaluation}
\label{sec:str2seq_exp}

In this section, we verify the sequence-structure alignment capability based on two experiments, i.e., \strtoseq\ prediction and zero-shot \seqtostr\ prediction via SSP and RNA design tasks, respectively. Overall, the results in \pref{tab:str2seq} and \pref{tab:seq2str} provide reliable evaluations of the FMs' capabilities in sequence-structure alignment. This underscores \our's efficacy in enabling genomic information to freely flow between structures and sequences.

\paragraph{RNA Design (\strtoseq) Evaluation}

we demonstrate the \strtoseq\ prediction capability of \our\ based on RNA design. We employed the Eterna~\citep{Lee14} V2 benchmark, which consists of $100$ specified secondary structures. This task aims to design RNA sequences based on reference structures. We develop a genetic algorithm (GA) which exploits masked nucleotide modelling (a.k.a., masked language modelling) to find plausible RNA sequences that solve RNA design puzzles. The implementation details can be found in \pref{fig:design} in \pref{app:rna_design}.
In the GA, the population size is set at $1000$, with $100$ iterations, and the mutation rate for each base is $0.5$. The evaluation metric is accuracy following existing works which indicates the number of puzzles solved by FMs. The experimental results are available in \pref{tab:str2seq}.

\begin{table}[htbp]
  \centering
\caption{Performance on the EternaV2 RNA design benchmark. The best accuracy is in \textbf{bold} face. ``Token.'' indicates the tokenization method.}
    \resizebox{.55\linewidth}{!}{
    \begin{tabular}{lcc}
    \toprule
    \textbf{Model} & \textbf{Token.} & \textbf{EternaV2 (Acc)}  \\
    \midrule
    RNAInverse & ---  & $30$ \\
    \midrule
    3UTRBERT  & k-mers   & $0$ \\
    DNABERT2  & BPE   & $0$ \\
    SpliceBERT  & SNT   & $3$ \\
    RNA-MSM  & SNT   & $2$ \\
    RNA-FM  & SNT   & $3$ \\
    \midrule
    \oursmallplus & SNT   & $71$ \\
    \ourbaseplus & SNT   & $\mathbf{74}$ \\
    \bottomrule
    \end{tabular}%
    }
  \label{tab:str2seq}%
  \vspace{-10pt}
\end{table}%

We include a popular baseline of RNAInverse and select recent DNA and RNA FMs which support masked language modelling. We exclude HyenaDNA in this experiment because it does not support masked nucleotide prediction. 
It is observed from \pref{tab:str2seq} that RNAInverse solved $30$ of the RNA design puzzles, indicating a promising capability in RNA design.
The FMs, such as 3UTRBERT and DNABERT2 fail in RNA design because they cannot handle SN-level modelling. Meanwhile, RNA-MSM, RNA-FM and SpliceBERT demonstrated trivial proficiency in RNA design, solving $2$ to $3$ puzzles. This observation suggests these FMs cannot precisely predict the bases without any \strtoseq\ prediction ability. With the help of \strtoseq, i.e., structure-contextualised sequence reconstruction, \oursmallplus\ and \ourbaseplus\ significantly outperformed other FMs with $71$ and $74$ puzzles solved, respectively, underscoring the significance of \strtoseq\ in sequence-structure alignment. Besides, we expect an increase in performance with sufficient computational budgets and the findings provide crucial evidence of the significance of \strtoseq\ for RNA sequence design.

\paragraph{Zero-shot SSP (\seqtostr) Evaluation}
\label{sec:seq2str_exp}
This subsection evaluates both \seqtostr\ and \strtoseq\ prediction in sequence-structure alignment. The evaluation of \seqtostr\ is based on zero-shot SSP. We use \our\ and \ourplus\ without fine-tuning to predict the secondary structures of sequences from the testing datasets and measure the macro-F1 score, where better structure prediction performance indicates a stronger capability for \seqtostr\ prediction. The experimental results are available in \pref{tab:seq2str}.

\begin{table}[htbp]
  \centering
  \caption{Performance in zero-shot SSP. The results are based on zero-shot inferences without any fine-tuning or domain adaptation. ``Stralign'' denotes the RNAStralign dataset.}
  \resizebox{.6\linewidth}{!}{
    \begin{tabular}{lccc}
    \toprule
    \multirow{2}[3]{*}{Model} & \multicolumn{3}{c}{\textbf{Zero-shot SSP (F1)}} \\
    \cmidrule(lr){2-4}
          & \textbf{Archive2} & \textbf{bpRNA} & \textbf{Stralign} \\
    \midrule
    ViennaRNA       & $73.99$ & $65.04$ & $74.09$ \\
    \midrule
    \oursmall       & $69.93$ & $65.85$ & $74.71$ \\
    \ourbase        & $74.38$ & $66.19$ & $74.91$ \\
    \oursmallplus   & $73.58$ & $65.95$ & $75.16$ \\
    \ourbaseplus    & $\mathbf{74.72}$  & $\mathbf{66.37}$ & $\mathbf{75.80}$ \\
    \bottomrule
    \end{tabular}%
  }
  \label{tab:seq2str}%
  \vspace{-10pt}
\end{table}

\begin{table*}[t!]
  \centering
  \caption{Performance of \our\ and baseline FMs on PGB. ``PolyA'' stands for Polyadenylation, ``Chrom Acc'' for Chromatin Accessibility, ``Prom Str'' for Promoter Strength, ``Term Str'' for Terminator Strength, ``Splice'' for Splice Site, ``Gene Exp'' for Gene Expression, and ``Enh Reg'' for Enhancer Region. Results for \ourbaseplus\ are excluded due to the time-intensive nature of the experiments.}
  \setlength{\tabcolsep}{2pt} 
  \resizebox{.6\linewidth}{!}{
    \begin{tabular}{lcccccccc}
    \toprule
    \multirow{2}[1]{*}{\textbf{Model}} & \textbf{PolyA} & \textbf{LncRNA} & \textbf{Chrom Acc} & \textbf{Prom Str} & \textbf{Term Str} & \textbf{Splice} & \textbf{Gene Exp} & \textbf{Enhancer} \\
    \cmidrule{2-9}    & F1 & F1 & F1 & RMSE & RMSE & F1 & RMSE & F1 \\
    \midrule
    DNABERT2  & $41.35$ & $72.55$ & $61.49$ & $0.99$ & $0.24$ & $45.34$ & $14.78$ & $36.40$ \\
    HyenaDNA  & $83.11$ & $58.21$ & $52.20$ & $0.88$ & $0.26$ & $90.28$ & $14.79$ & $66.17$ \\
    Caduceus  & $70.89$ & $68.40$ & $64.53$ & $0.91$ & $0.26$ & $78.51$ & $14.72$ & $60.83$ \\	
    NT-V2     & $71.26$ & $73.08$ & $65.71$ & $0.81$ & $0.27$ & $95.05$ & $14.79$ & $73.89$ \\
    Agro-NT   & $78.89$ & $67.24$ & $63.27$ & $0.94$ & $0.78$ & $88.45$ & $15.56$ & $62.83$ \\
    SpliceBERT& $65.23$ & $71.88$ & $63.62$ & $0.75$ & $0.22$ & $96.45$ & $14.70$ & $69.71$ \\
    3UTRBERT  & $76.48$ & $70.75$ & $63.71$ & $1.04$ & $0.36$ & $94.44$ & $14.87$ & $71.67$ \\
    RNA-BERT & $78.54$ & $61.99$ & $48.94$ & $1.81$  & $0.38$  & $94.45$ & $14.89$ & $57.61$ \\
    RNA-MSM & $84.25$ & $67.49$ & $53.52$ & $1.28$  & $0.28$  & $95.49$ & $14.87$ & $61.45$ \\
    RNA-FM & $84.94$ & $68.75$ & $54.92$ & $0.95$  & $0.27$  & $95.95$ & $14.83$ & $57.14$ \\
    \midrule
    \oursmall & $85.47$ & $75.71$ & $64.23$ & $0.67$ & $0.21$ & $97.40$ & $14.76$ & $68.31$ \\
    \ourbase  & $86.87$ & $77.53$ & $66.88$ & $0.65$ & $0.19$ & $98.15$ & $14.76$ & $72.45$ \\
    \midrule
    \oursmallplus & $87.05$ & $76.23$ & $65.41$ & $0.65$ & $0.20$ & $97.70$ & $14.76$ & $70.71$ \\
    \ourbaseplus & $\mathbf{87.55}$ & $\mathbf{77.96}$ & $\mathbf{67.69}$ & $\mathbf{0.59}$ & $\mathbf{0.18}$ & $\mathbf{98.41}$ & $\mathbf{14.71}$ & $\mathbf{79.77}$ \\
    \bottomrule
    \end{tabular}
  }
  \label{tab:pgb_results}
  \vspace{-10pt}
\end{table*}

The results in \pref{tab:seq2str} indicate that \our\ FMs mirrored the zero-shot secondary structure prediction (i.e., \seqtostr) performance of ViennaRNA. Moreover, \oursmallplus\ and \ourbaseplus\ outperform \our\ FMs based on structure contexts from ViennaRNA. Given the ablation of structure contexts, \ourbase\ also achieves performance comparable with ViennaRNA on the Archive2, bpRNA and RNAStralign datasets. Besides, we found that \ourplus\ generally obtains better performance on a wide genome downstream tasks owing to the structure awareness, and random or noise structure contexts have no obvious effects on the structure prediction. We cannot compare with other FMs in the zero-shot SSP experiments, because existing FMs were not pertained for secondary structure prediction.

\subsection{Results on RGB}
\label{sec:rgb_exp}
\begin{table}[H]
	\centering
	\caption{The performance of \our\ and baseline models on the RGB, with results averaged based on five random seeds. ``N.A.'' means not available for predictive tasks.}
        \setlength{\tabcolsep}{3pt} 
	\resizebox{\linewidth}{!}{
	\begin{tabular}{lcccccc}
		\toprule
		\multirow{2}[4]{*}{\textbf{Model}} & \textbf{mRNA}  &  \textbf{SNMD}  & \textbf{SNMR}  &  \textbf{Archive2} &  \textbf{Stralign} &  \textbf{bpRNA} \\
		\cmidrule{2-7}          & RMSE  & AUC   & F1    & F1    & F1    & F1 \\
		\midrule
		ViennaRNA & N.A.    & N.A.    & N.A.    & $73.99$ & $74.09$ & $65.03$ \\
		MXFold2     & N.A.    & N.A.    & N.A.    & $90.09$ & $97.01$ & $64.99$ \\
		Ufold           & N.A.    & N.A.    & N.A.    & $89.78$ & $95.76$ & $78.38$  \\
		\midrule
		DNABERT2 & $0.8158$ & $49.94$ & $15.86$ & $55.73$ & $64.09$ & $33.77$ \\
		HyenaDNA & $0.8056$ & $53.32$ & $39.80$  & $71.18$ & $91.24$ & $57.43$ \\
            Caduceus & $0.8026$ & $57.01$ & $39.59$ & $74.37$ & $92.28$ & $59.76$ \\
		NT-V2 & $0.7826$ & $50.49$ & $26.01$ & $68.36$ & $83.18$ & $56.95$ \\
		Agro-NT & $0.7830$ & $49.99$ & $26.38$ & $62.81$ & $72.54$ & $46.87$ \\
		SpliceBERT & $0.7340$ & $58.11$ & $46.44$ & $79.89$ & $93.81$ & $71.59$ \\
		3UTRBERT & $0.7772$ & $50.02$ & $24.01$ & $68.62$ & $88.55$ & $57.90$  \\
		RNABERT & $0.8087$ & $51.32$ & $29.14$ & $24.66$ & $83.68$ & $47.96$ \\
		RNA-MSM & $0.7321$ & $57.86$ & $45.22$ & $68.72$ & $91.15$ & $64.44$ \\
		RNA-FM & $0.7297$ & $59.02$ & $42.21$ & $82.55$ & $95.07$ & $78.16$ \\
		\midrule
		\oursmall & $0.7191$ & $62.44$ & $49.91$ & $88.48$ & $97.46$ & $80.51$ \\
		  \ourbase & $0.7164$ & $63.81$ & $50.80$  & $90.32$ & $97.82$ & $83.09$ \\
		\oursmallplus & $0.7174$ & $63.11$ & $51.21$ & $88.58$ & $97.33$ & $81.29$ \\
		\ourbaseplus & $\mathbf{0.7121}$ & $\mathbf{64.13}$ & $\mathbf{52.44}$ & $\mathbf{91.89}$ & $\mathbf{98.21}$ & $\mathbf{83.18}$ \\
		\bottomrule
	\end{tabular}%
	}
	\vspace{-10pt}
	\label{tab:rgb_results}%
\end{table}%

The results in \pref{tab:rgb_results} demonstrate the performance of \our\ and its generalizability across various fine-grained RNA downstream tasks. It is observed that \our\ models achieve better results than both RNA and DNA FM baselines, including Agro-NT and DNABERT2, which contain hundreds of millions of parameters. This is because the existing FMs usually adopt k-mers or BPE tokenization that cannot handle SN resolution tasks, e.g., single nucleotide mutation detection and repair, and structure prediction. Because of the \seqtostr\ pre-training, \our\ and \ourplus\ models exhibit strong results in secondary structure prediction, underscoring \our's capabilities in SN-level RNA sequence understanding and manipulation.

\subsection{Results on PGB}
\label{sec:pgb_exp}

The PGB is a plant-oriented genomic benchmark. Although the benchmark datasets in PGB are DNA-based tasks, we can still evaluate the performance of \our\ and its generalizability on multi-modal (i.e., DNA and RNA) genomic tasks.
The results in \pref{tab:pgb_results} reveal substantial variability in the performance of different FMs, where \oursmall\ outperformed other baseline models across most tasks, particularly in tasks like Polyadenylation, Splice Site, and Enhancer Region classification, where they achieved the highest F1 scores. This suggests that \our's architecture is particularly adept at handling complex genomic sequences. In comparison, existing FMs, e.g., NT-V2 and Agro-NT, showed lower performance with more parameters than \our. Besides, the performance of \oursmallplus\ suggests that the structure context can further enhance the performance of genomic modelling. 
Overall, \our\ models achieve state-of-the-art performance on both benchmarks, especially for \ourplus\ variants. The results underscore the importance of sequence-structure alignment in achieving complex genomic modelling tasks.

\section{Related Works}
\label{sec:related}
Current RNA FMs focused on sequence-to-structure mapping, e.g., end-to-end secondary structure prediction. However, to the best of our knowledge, the sequence-structure alignment in RNA genome modelling has yet been investigated in the literature. 
There have been some preliminary works, such as scBERT~\citep{YangWWFTHLY22}, RNABERT~\citep{AkiyamaS22}, RNA-FM~\citep{Chen22}, RNA-MSM~\citep{Zhang23}, and RNAErnie~\citep{WangBLLMKX24}, to name a few. However, these methods have only trained the FMs on a limited-scale database, as RNA sequences are generally expensive to obtain. Some FMs focus on specific types of RNA sequences, such as coding sequences (CDS)~\citep{HalleeRG23}, 5' untranslated regions (5'UTR)~\citep{Chu24}, 3' untranslated regions (3'UTR)~\citep{YangLP23}, or precursor mRNA sequences~\citep{ChenZD23}, thus limiting the models' ability to capture the diversity of RNA sequences. Uni-RNA~\citep{WangGCLJKW23} has been reported to achieve good performance due to the large scale of the model and database, however, it is not open-sourced and cannot be compared in the experiments.

In short, the existing RNA FMs neglect the significance of sequence-structure alignment in RNA genome modelling, while the 5UTR-LM~\citep{Chu24} adopts the secondary structure prediction as a pre-training objective to achieve \seqtostr\ prediction in pre-training. However, these FMs are not available for \strtoseq\ mapping and suffer from limited model and data scales that fail to uncover the comprehensive efficacy of sequence-structure alignment on a wide set of genomic tasks. ERNIE-RNA~\citep{Yin24} feeds the RNA structure along with the sequence into the model and improves the downstream tasks. However, it also ignores the significance of \strtoseq\ prediction capability. In a nutshell, existing FMs fail to achieve sequence-structure alignment without exception.

\section{Conclusion}

We introduced \our\ to tackle the challenge of sequence-structure alignment in genome modelling, which bridges the gap between sequence and structural information and improves the reliability of genome analysis.
Experimental results on four comprehensive in-silico RNA and DNA benchmarks demonstrate that \our\ outperforms existing FMs across diversified downstream tasks, e.g., up to $98\%$ F1 score for SSP and $74\%$ accuracy of RNA design. The superior performance highlights the potential of sequence-structure alignment in the field of genomics.

\section*{Acknowledgements}
This work was supported in part by the UKRI Future Leaders Fellowship under Grant MR/S017062/1 and MR/X011135/1; in part by NSFC under Grant 62376056 and 62076056; in part by the Royal Society under Grant IES/R2/212077; in part by the EPSRC under Grant 2404317; in part by the Kan Tong Po Fellowship (KTP\textbackslash R1\textbackslash 231017); and in part by the Amazon Research Award and Alan Turing Fellowship.

\appendix
\bibliography{ref}

\section{Reproducibility Checklist}
This paper:
\begin{itemize}
    \item Includes a conceptual outline and/or pseudocode description of AI methods introduced \textbf{yes}
    \item Clearly delineates statements that are opinions, hypothesis, and speculation from objective facts and results \textbf{yes}
    \item Provides well marked pedagogical references for less-familiare readers to gain background necessary to replicate the paper \textbf{yes}
\end{itemize}

\textbf{Does this paper make theoretical contributions?} \textbf{yes}

If yes, please complete the list below.
\begin{itemize}
    \item All assumptions and restrictions are stated clearly and formally. \textbf{yes}
    \item All novel claims are stated formally (e.g., in theorem statements). \textbf{yes}
    \item Proofs of all novel claims are included. \textbf{yes}
    \item Proof sketches or intuitions are given for complex and/or novel results. \textbf{yes}
    \item Appropriate citations to theoretical tools used are given. \textbf{yes}
    \item All theoretical claims are demonstrated empirically to hold. \textbf{yes}
    \item All experimental code used to eliminate or disprove claims is included. \textbf{yes}

\end{itemize}

\textbf{Does this paper rely on one or more datasets?} \textbf{yes}

If yes, please complete the list below.

\begin{itemize}
    \item A motivation is given for why the experiments are conducted on the selected datasets \textbf{yes}
    \item All novel datasets introduced in this paper are included in a data appendix. \textbf{yes}
    \item All novel datasets introduced in this paper will be made publicly available upon publication of the paper with a license that allows free usage for research purposes. \textbf{yes}
    \item All datasets drawn from the existing literature (potentially including authors’ own previously published work) are accompanied by appropriate citations. \textbf{yes}
    \item All datasets drawn from the existing literature (potentially including authors’ own previously published work) are publicly available. \textbf{yes}
    \item All datasets that are not publicly available are described in detail, with explanation why publicly available alternatives are not scientifically satisficing. \textbf{yes}
\end{itemize}

\textbf{Does this paper include computational experiments?} \textbf{yes}

If yes, please complete the list below.

\begin{itemize}
    \item Any code required for pre-processing data is included in the appendix. \textbf{yes}
    \item All source code required for conducting and analyzing the experiments is included in a code appendix. \textbf{yes}
    \item All source code required for conducting and analyzing the experiments will be made publicly available upon publication of the paper with a license that allows free usage for research purposes. \textbf{yes}
    \item All source code implementing new methods have comments detailing the implementation, with references to the paper where each step comes from \textbf{yes}
    \item If an algorithm depends on randomness, then the method used for setting seeds is described in a way sufficient to allow replication of results. \textbf{yes}
    \item This paper specifies the computing infrastructure used for running experiments (hardware and software), including GPU/CPU models; amount of memory; operating system; names and versions of relevant software libraries and frameworks. \textbf{yes}
    \item This paper formally describes evaluation metrics used and explains the motivation for choosing these metrics. \textbf{yes}
    \item This paper states the number of algorithm runs used to compute each reported result. \textbf{yes}
    \item Analysis of experiments goes beyond single-dimensional summaries of performance (e.g., average; median) to include measures of variation, confidence, or other distributional information. \textbf{yes}
    \item The significance of any improvement or decrease in performance is judged using appropriate statistical tests (e.g., Wilcoxon signed-rank). \textbf{N.A. because of computation complexity }
    \item This paper lists all final (hyper-)parameters used for each model/algorithm in the paper’s experiments. \textbf{yes}
    \item This paper states the number and range of values tried per (hyper-) parameter during development of the paper, along with the criterion used for selecting the final parameter setting. \textbf{yes}
\end{itemize}

\section{DNA Foundation Models}

Biological sequence modelling, including DNA, RNA, and protein, has attracted attention in recent years. Protein modelling, e.g., AlphaFold~\citep{Jumper21,Evans21,Abramson24} and ESM~\citep{Lin22}, has been studied for many years compared to DNA and RNA modelling. In the realm of genomic sequence modelling, several early works aimed at addressing diversified genome downstream subtasks. For instance, DNABERT~\citep{JiZLD21} adapts the architecture of BERT~\citep{DevlinCLT19} for genomic sequence modelling, showing preliminary performance for in-silico genomic tasks. DNABERT2~\citep{Zhou23}, a multi-species FM improved based on DNABERT, proposes replacing k-mers tokenization with BPE tokenization to improve model performance. To explore the performance of large-scale FMs, the nucleotide transformers (V1 \& V2)~\citep{Dalla23}, AgroNT~\citep{Mendoza23} and SegmentNT~\citep{De24} leveraged billions of parameters to boost genomic sequence modelling and achieved promising performance in understanding DNA genome, with model scales up to $2.5$ billion and $1$ billion parameters, respectively. Agro-NT~\citep{Mendoza23} was pre-trained on multi-species edible plant DNA sequences but failed to transfer effectively to RNA sequence modelling in our experiments. To address the modelling capacity problem caused by the remarkable lengths of genomes, there is a growing focus on the necessity of long-range sequence modelling and the introduction of autoregressive FMs, namely, HyenaDNA~\citep{Nguyen23} and Evo~\citep{Nguyen24}.

\section{Pre-training Environment}
\label{app:environment}

The pre-training of \our{} was conducted on a dedicated Linux computation node, equipped with $8$ \textsc{Nvidia} RTX $4090$ GPUs. For distributed model training, we employed version $4.37.1$ of the Transformers library alongside version $0.26.1$ of the Accelerate library. Our implementation framework of choice for \our\ was PyTorch, specifically version $2.0.0$. The ViennaRNA version is $2.6.4$ in our experiments. While some existing code was adapted for the modules within \our, the majority of the codebase, such as genomic sequences preprocessing, model pre-training, objective functions, and experiments, was meticulously crafted from scratch.

\section{\our~Package}
\label{app:package}

Genome modelling is still in its early stages. Therefore, open-source codes and resources are consequently very scarce. Typically, existing FMs usually release the pre-trained model weights only, without providing the training and fine-tuning codes, and benchmark evaluation scripts, etc. 

To address this issue, we have developed a comprehensive open-source RNA genome modelling toolkit\footnote{\url{https://github.com/yangheng95/OmniGenBench}} based on \our. This toolkit aims to provide extensive FM fine-tuning tutorials and a unified automated benchmark evaluation. The main features of the \our\ Package are as follows:
\begin{itemize}
    \item \textbf{Fine-Tuning Tutorials}: We provide tutorials for fine-tuning on all downstream genome modelling tasks, including dataset processing, model implementation, and training processes. A fine-tuning example for secondary structure is included, covering both training and demonstration of secondary structure prediction. 
    \item \textbf{Automated Benchmark Evaluation}: We offer an automated benchmark evaluation interface, which includes the built-in PGB and RGB benchmarks. By predefining the configurations for benchmark evaluation subtasks, such as hyperparameters, our tool supports the automated benchmark evaluation of future FMs and the addition of new benchmarks. The goal of automated benchmark evaluation is to ensure fairness and ease of use. We provide a tutorial on automated evaluation to guide users in benchmark evaluation. The automated benchmarking tutorial is available at: \url{https://github.com/yangheng95/OmniGenBench/blob/master/examples/benchmark/autobench.ipynb}.
    \item \textbf{Online Hub for Genome Modelling}: We have created a hub for hosting and distributing open-source licensed datasets, model checkpoints, and benchmark evaluations. Additionally, we have designed flexible interfaces to support the sharing of datasets and models within the community. This approach helps mitigate the issue of resource scarcity. The hub will be available soon.
\end{itemize}

We are in the process of finalising the necessary documentation and will officially release this tool in the near future.

\begin{table*}[t!]
  \centering
  \caption{The brief statistics of RNA and DNA FM baselines. Please note that the pertaining data scales cannot be directly compared because the measurements are different in various publications. The detailed introduction of these FMs can be found in original publications. }
  \resizebox{\linewidth}{!}{
    \begin{tabular}{lccccccc}
    \toprule
    \textbf{Model} & \textbf{Tokenization} & \textbf{\# of Params} & \textbf{Pre-training Data Scale} & \textbf{Pre-training Data Source} & \textbf{Species} & \textbf{Sequence Type} \\
    \midrule
    DNABERT-2 & BPE & $117$M & $32.49$B Tokens & The 1000 Genomes Project & Human + $135$ Species & DNA \\
    NT-V2-$100$M & k-mers & $96$M & $300$B Tokens & The 1000 Genomes Project, etc. & Human + $850$ Species & DNA \\
    HyenaDNA-Large & SNT & $47$M & $3.2$B Tokens & Genome Reference Consortium & Human & DNA \\
    Caduceus     & SNT & $1.9$M  & $35$B Tokens  & Genome Reference Consortium & Human & DNA \\
    Agro-NT-$1$B & k-mers & $985$M & $472.5$B Tokens & Ensembl Plants Database & $48$ Edible Plants & DNA \\
    \midrule
    SpliceBERT & SNT & $19$M & $2$M Sequences & UCSC Genome Browser & Multi-Vertebrates & precursor-mRNA \\
    RNA-BERT & SNT & $0.5$M & $4,069$ RNA Families & The RNA Families Database & Multi-Species & ncRNA \\
    RNA-MSM & SNT & $96$M & $4,069$ RNA Families & The RNA Families Database & Multi-Species & ncRNA \\
    RNA-FM & SNT & $96$M & $23$M Sequences & RNAcentral Database & Multi-Species & ncRNA \\
    3UTRBERT & k-mers & $86$M & $20,362$ Sequences & The GENCODE Project & Human & mRNA $3$'UTR \\
    \midrule
    \oursmall & \multirow{2}[1]{*}{SNT} & $52$M & \multirow{2}[1]{*}{$54.2$B Tokens} & \multirow{2}[1]{*}{The OneKP Initiative} & \multirow{2}[1]{*}{$1124$ Plant Species} & \multirow{2}[1]{*}{mRNA, CDS, UTR} \\
    \ourbase & & $186$M & & & & \\
    \bottomrule
    \end{tabular}%
    }
  \label{tab:fm_details}%
  \vspace{-10pt}
\end{table*}%

\section{Comparison Baselines}
\label{app:baselines}
There are no direct counterparts to \our\ aimed at plant RNA genome modelling. We can compare it with the following recent RNA and DNA genome FMs as potential baselines to help evaluate the performance of \our. The brief introductions of the FMs in \pref{tab:fm_details} are as follows:

\begin{itemize}
    \item ViennaRNA~\citep{Lorenz11}. ViennaRNA is a comprehensive genomic analysis tool that includes a diverse set of interfaces, such as RNAFold\footnote{\url{https://www.tbi.univie.ac.at/RNA/RNAfold.1.html}} and RNAInverse\footnote{\url{https://www.tbi.univie.ac.at/RNA/RNAinverse.1.html}} design. ViennaRNA serves as the baseline for RNA structure prediction and RNA design in our experiments.
    \item MXFold2~\cite{Sato21MXFold2}. MXFold2 is a deep learning model for RNA secondary structure prediction, integrating thermodynamic models to improve prediction accuracy. It stands out for its robustness across various RNA families and is widely used for predicting RNA secondary structures from given sequences.
    \item UFold~\cite{Fu22UFold}. UFold is a deep learning-based RNA secondary structure prediction tool designed for speed and accuracy. It uses a convolutional neural network (CNN) architecture optimized to predict base-pairing probabilities in RNA sequences, enabling efficient secondary structure prediction.
    \item DNABERT2~\citep{Zhou23}. DNABERT2 is one of the latest DNA FMs which improves the performance of DNABERT. The main modification of DNABERT2 is the tokenization method, which was changed to BPE from k-mers.
    \item HyenaDNA~\citep{Nguyen23}. HyenaDNA is an autoregressive FM optimised for long-range genome data processing. HyenaDNA is based on the Hyena convolution architecture and capable of handling sequences up to $1$M bases in length.
    \item Caduceus~\cite{Schiff24caduceus}. Caduceus\footnote{\url{https://huggingface.co/kuleshov-group/caduceus-ps_seqlen-131k_d_model-256_n_layer-16}} is an advanced DNA language model built on the MambaDNA architecture, designed to address challenges in genomic sequence modelling, such as long-range token interactions and reverse complementarity (RC). 
    \item Nucleotide Transformer (NT) V2~\citep{Dalla23}. The NT FMs were trained on DNA data, including the human reference genome and multi-species DNA sequences. They aim to capture the complex patterns within nucleotide sequences for various genome modelling applications.
    \item Agricultural Nucleotide Transformer (Agro-NT)~\citep{Mendoza23}. Agro-NT is a large-scale DNA FM ($1$B parameters) akin to the Nucleotide Transformers but with a focus on plant DNA.
    \item SpliceBERT~\citep{ChenZD23}. It was trained on $2$M precursor messenger RNA (pre-mRNA) and specialized in RNA splicing of pre-mRNA sequences.
    \item 3UTRBERT~\citep{YangLP23}. This model was trained on $20$k 3'UTRs for 3'UTR-mediated gene regulation tasks. It uses k-mers tokenization instead of SNT.
    RNA-BERT~\cite{AkiyamaS22}. RNA-BERT is a BERT-style model pre-trained on a large corpus of non-coding RNA sequences. It uses masked language modelling (MLM) as its primary training objective. The model is designed to predict RNA structural alignments and can be fine-tuned for various RNA sequence classification and regression tasks
    \item RNA-MSM~\cite{ZhangLJGXLCSHXS24} RNA-MSM is an unsupervised RNA language model based on multiple sequence alignment (MSA). It is the first model of its kind to produce embeddings and attention maps that directly correlate with RNA secondary structure and solvent accessibility. RNA-MSM is particularly effective for tasks involving evolutionary relationships in RNA sequences.
    \item RNA-FM~\cite{Chen22} RNA-FM is a BERT-based RNA foundation model trained on a vast dataset of non-coding RNA sequences. The model excels in predicting RNA structure and function by leveraging masked language modelling (MLM) during pre-training. RNA-FM's training data is sourced from the RNAcentral database, providing it with extensive knowledge across diverse RNA species.
    \item \our. \our\ is the RNA genome FM that advocates the importance of sequence-structure alignment. Moreover, it is the first FM which addressed the in-silico RNA design task.
    \item \ourplus\footnote{Please find the finetuning example\footnote{\url{https://anonymous.4open.science/r/OmniGenome-SupplementaryMaterials-002F/tutorials/ssp_finetuning.py}} of \ourplus\ in the supplemental materials.}. \ourplus\ is a variant of \our\ that feeds both sequences and structures into \our\ to aggregate the feature representations to improve modelling ability. 
\end{itemize}
These genomic FMs serve as valuable baselines for evaluating the performance of \our\ in our study, particularly in RNA structure prediction and sequence-to-structure mapping tasks.

\section{Benchmark Suites}
\label{sec:benchmarks}
In this section, we first introduce the details, such as downstream task types and species, of RGB and PGB. Besides, we have included two recent DNA benchmarks in the evaluation, aiming to provide comprehensive performance comparisons between \our\ with existing DNA FMs, where \our\ was not pre-trained on any specific DNA genome database.

\subsection{RNA Genomic Benchmark (RGB)}
\label{app:rgb}
RGB contains $7$ SN-level tasks that are curated or collected from the literature, and the details of the RGB can be found in \pref{tab:rgb_detail}. The objective of RGB is to benchmark genome FMs in challenging SN-level modelling tasks such as the detection and repair of SN mutations, mRNA sequence degradation rates, and RNA secondary structure prediction. Due to the lack of a plant RNA benchmark dataset, RGB enriches the benchmark suites by including the modelling of RNA downstream tasks from a variety of species, e.g., plants and animals. The sequence length in RGB ranges from $107$ to $512$, which is sufficient for most RNA understanding tasks. In summary, these multi-species and SN-level tasks in RGB serve as the first comprehensive benchmark utilised to assess the RNA sequence modelling capabilities of \our\ and its baseline models. The brief introduction of the datasets in RGB is as follows:

\begin{itemize}[leftmargin=*]

\item \textbf{Single-Nucleotide Mutation Detection (SNMD)}: We developed a plant RNA dataset synthesising the single-nucleotide mutations. Focused on identifying potential single nucleotide changes, this task is essential for detecting mutations linked to genetic disorders. The SNMD dataset introduces up to $10$ random mutations in the original sequences, regardless of variation ratios. Cross-entropy is utilised as the loss function for this binary token classification task.

\item \textbf{Single-Nucleotide Mutation Repair (SNMR)}: This task challenges the model to suggest corrective actions at the single nucleotide level, aiding in gene therapy approaches. The SNMR dataset mirrors the SNMD dataset, with cross-entropy as the loss function, indicating a token 4-way (i.e., \texttt{A}, \texttt{U}, \texttt{C}, \texttt{G}) classification task.

\item \textbf{mRNA Degrade Rate Prediction (mRNA)}: Estimating the decay rate of nucleotide bases in mRNA sequences, this task is vital for deciphering gene expression and regulation. The dataset originates from the Kaggle COVID-19 vaccine design competition\footnote{\url{https://www.kaggle.com/competitions/stanford-covid-vaccine}}, focusing solely on sequence-based degradation rate prediction and excluding RNA structures. It's a token regression task using MSE as the loss function, with the dataset re-split into training, validation, and testing sets for evaluation.

\item \textbf{RNA Secondary Structure Prediction (bpRNA \& Archive2 \& RNAStralign)}: Aiming to predict RNA folding into secondary structures, this task is fundamental to RNA functionality and interactions. We evaluated \our\ on four datasets, bpRNA~\citep{DanaeeRWDHH18}, ArchiveII~\citep{Mathews19}, RNAStralign~\citep{TanFSM17} and Rfam~\citep{Kalvari21}. Following existing works, we have excluded sequences over $512$ bases and complex structures, simplifying to three symbols: \texttt{`('}, \texttt{`.'}, \texttt{`)'}\. We have filtered the RNAStralign, Archive2 and bpRNA datasets using CD-HIT-EST and BLAST as described in \pref{sec:filtering}, which results in different data splits compared to existing works for SSP. Please find the dataset details of splits in \pref{tab:rgb_detail}. Besides, the SSP datasets processed in different publications are usually unknown because of various data filtering implementations. As a result, the performance of \our\ may not be directly compared with other studies. RNA SSP tasks are trained based on cross-entropy loss functions.

\end{itemize}

\begin{table*}[t!]
  \centering
  \caption{The brief statistics of subtasks in the RGB. These benchmark datasets are held out or not included in the pre-training database. The numbers of examples in training, validation and testing sets are separated by ``/''. ``StrAlign'' indicates the RNAStrAlign dataset.}
  \resizebox{.9\linewidth}{!}{
    \begin{tabular}{lcccccc}
    \toprule
     \textbf{Task} & \multicolumn{1}{c}{\textbf{Task Type}} & \multicolumn{1}{c}{\textbf{\# of examples}} & \multicolumn{1}{c}{\textbf{\# of classes}} & \multicolumn{1}{c}{\textbf{Metric}} & \multicolumn{1}{c}{\textbf{Sequence length}} & \multicolumn{1}{c}{\textbf{Source}} \\
    \midrule
    SNMD & Token classification & $8,000/1,000/1,000$ & $2$ & AUC & $200$ & This work \\
    SNMR & Token classification & $8,000/1,000/1,000$ & $4$ & F1 & $200$ & This work \\
    mRNA & Token regression & $1,735/193/192$ &  ---  & RMSE & $107$ & Kaggle\footnote{\url{https://www.kaggle.com/competitions/stanford-covid-vaccine}} \\
    bpRNA & Token classification & $9,232/1,154/1,161$ & $3$ & F1 & $\leq500$ & \cite{Franke24RNformer} \\
    AchiveII & Token classification & $608/76/82$ & $3$ & F1 & $\leq500$ & \citep{Mathews19} \\
    StrAlign & Token classification & $3104/389/388$ & $3$ & F1 & $\leq500$ & \citep{TanFSM17} \\
    \bottomrule
    \end{tabular}%
    }
    \label{tab:rgb_detail}%

\end{table*}%

Please find the appendix for the input and output examples of each subtask in RGB. The detailed task descriptions for each nucleic acid and species, including the number of examples, classes, evaluation metric, and sequence length, are outlined in \pref{tab:rgb_detail}. Each task is carefully curated to reflect the complexity and variety inherent in genomic data, providing a robust framework for assessing the nuanced capabilities of state-of-the-art RNA FMs.

\pref{tab:examples} show the virtual examples of different datasets in RGB. Please refer to our supplementary materials to find the datasets for more details.

\begin{table*}[htbp]
  \centering
  \caption{The virtual input and output examples in RGB. The ``$\dots$'' represents the sequences that are omitted for better presentation and the \textcolor{red}{red} colour indicates the wrong prediction in classification tasks. In the mRNA dataset, all single nucleotide bases have three values to predict. Note that ``\texttt{T}'' and ``\texttt{U}'' can be regarded as the same symbol in RNA sequences and depend on different datasets. }
  \resizebox{.7\linewidth}{!}{
    \begin{tabular}{cccc}
    \textbf{Genome~Type} & \textbf{Dataset}  &  & \textbf{Examples} \\
    \midrule
          \multirow{24}[1]{*}{RNA} & \multirow{3}[2]{*}{SNMD} & Input~Sequence & G~A~G~T~A~$\dots$~T~T~G~A~G \\
          &       & True~Label & 0~~0~~1~~0~~0~$\dots$~0~~0~~1~~0~~0 \\
          &       & Prediction & 0~~0~~\textcolor{red}{0}~~0~~0~$\dots$~0~~0~~1~~0~~0 \\
\cmidrule{2-4}          & \multirow{3}[2]{*}{SNMR} & Input~Sequence & T~A~C~G~A~~$\dots$~C~T~G~A~T \\
          &       & True~Label & T~A~C~A~A~$\dots$~G~T~A~A~T \\
          &       & Prediction & T~A~C~A~A~$\dots$~\textcolor{red}{C}~T~G~A~T \\
          
\cmidrule{2-4}    & \multirow{3}[2]{*}{mRNA} & Input~Sequence & G~G~$\dots$~A~C \\
          &       & True~Label & [0.1,0.3,0.2]~[0.8,0.4,0.1]$\dots$[0.9,0.4,0.3]~[0.5,0.2,0.6] \\
          &       & Prediction & [0.1,0.3,0.2]~[0.8,0.4,0.1]$\dots$[0.9,0.4,0.3]~[0.5,0.2,0.6] \\
\cmidrule{2-4}          & \multirow{3}[2]{*}{bpRNA} & Input~Sequence & G~G~C~G~A~$\dots$~C~U~U~U~U \\
          &       & True~Label & (~~~(~~~(~~~$\cdot$~~~$\cdot$~$\dots$~$\cdot$~~~$\cdot$~~~)~~~)~~~) \\
          &       & Prediction & (~~~(~~~(~~~\textcolor{red}{(}~~~$\cdot$~$\dots$~$\cdot$~~~\textcolor{red}{)}~~~)~~~)~~~) \\
\cmidrule{2-4}          & \multirow{3}[2]{*}{Archive2} & Input~Sequence & A~G~U~A~G~$\dots$~U~U~U~G~C~U \\
   &       & True~Label & (~~~(~~~(~~~$\cdot$~~~$\cdot$~~~$\dots$~$\cdot$~~~$\cdot$~~~)~~~)~~~) \\
          &       & Prediction & (~~~(~~~(~~~$\cdot$~~~$\cdot$~~~$\dots$~$\cdot$~~~$\cdot$~~~)~~~)~~~) \\
          \cmidrule{2-4}          & \multirow{3}[2]{*}{RNAStralign} & Input~Sequence & A~G~U~A~G~$\dots$~U~U~U~G~C~U \\
   &       & True~Label & (~~~(~~~(~~~$\cdot$~~~$\cdot$~~~$\dots$~$\cdot$~~~$\cdot$~~~)~~~)~~~) \\
          &       & Prediction & (~~~(~~~(~~~$\cdot$~~~$\cdot$~~~$\dots$~$\cdot$~~~$\cdot$~~~)~~~)~~~) \\
          \cmidrule{2-4}          & \multirow{3}[2]{*}{Rfam} & Input~Sequence & A~G~U~A~G~$\dots$~U~U~U~G~C~U \\
   &       & True~Label & (~~~(~~~(~~~$\cdot$~~~$\cdot$~~~$\dots$~$\cdot$~~~$\cdot$~~~)~~~)~~~) \\
          &       & Prediction & (~~~(~~~(~~~$\cdot$~~~$\cdot$~~~$\dots$~$\cdot$~~~$\cdot$~~~)~~~)~~~) \\
    \bottomrule
    \end{tabular}%
    }
   \label{tab:examples}%
\end{table*}%

\subsection{Plant Genomic Benchmark (PGB)}
\label{app:pgb}
The Plant Genomic Benchmark~\citep{Mendoza23} (PGB) provides a comprehensive suite of DNA downstream datasets designed to evaluate and improve the predictive capabilities of genomic FMs in plant biology. This benchmark, as shown in \pref{tab:pgb_detail}, encompasses a range of critical genomic tasks\footnote{https://huggingface.co/InstaDeepAI/agro-nucleotide-transformer-1b}, including binary classification, single and multi-variable regression, and multi-label classification, addressing various aspects of plant genomics such as RNA processing, gene expression, and chromatin accessibility. By integrating diverse genomic tasks, the PGB aims to facilitate advanced research and development in plant genomics, offering a robust platform for the assessment and enhancement of model performance across different plant species. To obtain a detailed description of PGB, please refer to Agro-NT~\citep{Mendoza23}.

\begin{table*}[t!]
    \centering
    \caption{The genomic tasks in the Plant Genomic Benchmark. This table briefly enumerates each task by name, the number of datasets available, the type of classification or regression analysis required, the range of sequence lengths, and the total number of samples in each dataset. Please find the dataset details of PGB in Agro-NT~\citep{Mendoza23}.}
    \resizebox{\linewidth}{!}{
        \begin{tabular}{lcccccc}
            \toprule
             \textbf{Task}  & \multicolumn{1}{c}{\textbf{\# of datasets}} & \multicolumn{1}{c}{\textbf{Task Type}} & \multicolumn{1}{c}{\textbf{Total \# of examples}} & \multicolumn{1}{c}{\textbf{\# of classes}} & \multicolumn{1}{c}{\textbf{Metric}} & \multicolumn{1}{c}{\textbf{Sequence length}} \\
            \midrule
            Polyadenylation & $6$ & Sequence classification & $738,918$  & $2$ & F1 & $400$  \\
            Splice site  & $2$  & Sequence classification & $4,920,835$  & $2$ & F1 & $398$  \\
            LncRNA &  $2$   & Sequence classification & $58,062$ & $6$ & F1 & $101-6000$  \\
            Promoter strength & $2$  & Sequence regression & $147,966$   & --- & RMSE & $170$  \\
            Terminator strength & $2$  & Sequence regression & $106,818$   & --- & RMSE & $170$  \\
            Chromatin accessibility & $7$   & Multi-label classification & $5,149,696$  & $9-19$ & F1 & $1,000$  \\
            Gene expression & $6$  & Multi-variable regression & $206,358$   & --- & RMSE & $6,000$  \\
            Enhancer region & $1$  & Sequence classification & $18,893$  &  $2$ & F1 & $1,000$  \\
            \bottomrule
        \end{tabular}
    }
    \label{tab:pgb_detail}
 
\end{table*}

\subsection{Genomic Benchmarks}
The genomic benchmark (GB) is also a DNA-oriented FM benchmark suite, which can be used for generalisability evaluation of \ourbase. It contains a well-curated collection of datasets designed for the classification of genomic sequences, focusing on regulatory elements across multiple model organisms. This collection facilitates robust comparative analysis and development of genomic FMs. The task names in the original repository are complex, we abbreviate the names as follows: 
\begin{itemize}
    \item DEM corresponds to ``Demo Coding vs Intergenomic Seqs''
    \item DOW is for ``Demo Human or Worm''
    \item DRE represents ``Drosophila Enhancers Stark''
    \item HCE is short for ``Human Enhancers Cohn''
    \item HEE denotes ``Human Enhancers Ensembl''
    \item HRE abbreviates ``Human Ensembl Regulatory''
    \item HNP shortens ``Human Nontata Promoters''
    \item HOR is an abbreviation for ``Human Ocr Ensembl''
    \item DME simplifies ``Dummy Mouse Enhancers Ensembl''
\end{itemize}
The brief statistics for each dataset included in the GUE benchmark are displayed in \pref{tab:gue_detail}.
Similar to GUE, we run the evaluation on a subset of GB, where for each task we randomly select at most $10$k samples from the original splits, e.g., training, testing and validation (if any) sets.

\begin{table*}[htbp]
    \centering
    \caption{The brief statistics of datasets reported in the genomic benchmark~\cite{Grevsova23}.}
    \resizebox{.7\linewidth}{!}{
    \begin{tabular}{lcccccc}
        \toprule
        \textbf{Task} & \textbf{\# of Sequences} & \textbf{\# of Classes} & \textbf{Class Ratio} & \textbf{Median Length} & \textbf{Standard Deviation} \\
        \midrule
        DME & $1,210$ & $2$ & $1.0$ & $2,381$ & $984.4$ \\
        DEM & $100,000$ & $2$ & $1.0$ & $200$ & $0.0$ \\
        DOW & $100,000$ & $2$ & $1.0$ & $200$ & $0.0$ \\
        DRE & $6,914$ & $2$ & $1.0$ & $2,142$ & $285.5$ \\
        HCE & $27,791$ & $2$ & $1.0$ & $500$ & $0.0$ \\
        HEE & $154,842$ & $2$ & $1.0$ & $269$ & $122.6$ \\
        HRE & $289,061$ & $3$ & $1.2$ & $401$ & $184.3$ \\
        HNP & $36,131$ & $2$ & $1.2$ & $251$ & $0.0$ \\
        HOR & $174,456$ & $2$ & $1.0$ & $315$ & $108.1$ \\
        \bottomrule
    \end{tabular}
    }
    \label{tab:gb_detail}
\end{table*}

\begin{table*}[htbp]
  \centering
  \caption{Performance of \our\ and baseline FMs across different tasks in the genomic benchmarks (GB), where the results are re-implemented based on our evaluation protocol. The performance (macro F1) for each task is the average macro F1 score in all sub-datasets.}
  \resizebox{.7\linewidth}{!}{
    \begin{tabular}{lccccccccc}
    \toprule
    \textbf{Model} & \textbf{DEM} & \textbf{DOW} & \textbf{DRE} & \textbf{DME} & \textbf{HCE} & \textbf{HEE} & \textbf{HRE} & \textbf{HNP} & \textbf{HOR} \\
    \cmidrule{2-10}      & F1 & F1 & \textbf{F1} & F1 & \textbf{F1} & F1 & F1 & F1 & F1 \\
    \midrule
    DNABERT-2 & $92.67$ & $95.17$ & $43.77$ & $77.21$ & $\mathbf{75.58}$ & $80.66$ & $78.14$ & $85.80$  & $68.03$ \\
    HyenaDNA & $88.21$ & $94.13$ & $70.11$ & $76.44$ & $70.38$ & $79.58$ & $96.33$ & $85.99$ & $67.03$ \\
    Caduceus & $92.13$ & $94.74$ & $72.03$ & $75.61$ & $70.20$ & $76.47$ & $79.16$ & $84.36$ & $63.17$ \\
    NT-V2  & $91.66$ & $94.32$ & $\mathbf{78.20}$  & $\mathbf{81.72}$ & $71.98$ & $79.85$ & $93.30$  & $85.30$  & $68.53$ \\
    SpliceBERT & $\mathbf{94.72}$ & $\mathbf{96.42}$ & $72.29$ & $74.70$  & $73.50$  & $79.60$  & $95.23$ & $\mathbf{89.57}$ & $68.89$ \\
    3UTRBERT  & $89.50$  & $90.22$ & $74.35$ & $80.14$ & $70.23$ & $76.33$ & $\mathbf{98.47}$ & $82.49$ & $66.78$ \\
    \ourbase & $94.16$ & $93.49$ & $77.17$ & $80.34$ & $73.51$ & $\mathbf{82.23}$ & $95.66$ & $87.87$ & $\mathbf{68.97}$ \\
    \bottomrule
    \end{tabular}%
    }
  \label{tab:gb_results}%
\end{table*}%

The experimental results presented in \pref{tab:gb_results} demonstrate that \ourbase\ consistently achieves competitive performance across a diverse array of genomic tasks. Notably, \ourbase\ excels in the Human Ensembl Regulatory (HRE) task with an F1 score of $95.66$, outperforming other models like DNABERT-2 and HyenaDNA in this specific benchmark. Additionally, \ourbase\ shows robust results in tasks involving enhancer predictions (HEE) and non-TATA promoters (HNP), underscoring its versatility and effectiveness in processing complex genomic sequences. These findings highlight the advanced capabilities of \ourbase\ in handling intricate genomic data, contributing significantly to the field of genomic research.

\subsection{Genomic Understanding Evaluation}
The Genome Understanding Evaluation~\cite{Zhou23} serves as a DNA genomic benchmark, encompassing $36$ datasets across nine crucial genome analysis tasks applicable to a variety of species. Similar to PGB and GB, it is used for evaluating the generalisability of \our\ on DNA genome benchmarking. To thoroughly assess the capabilities of genome foundation models across sequences of varying lengths, tasks have been chosen with input lengths spanning from $70$ to $10,000$. The brief statistics for each dataset included in the GUE benchmark are displayed in \pref{tab:gue_detail}, and the task descriptions are available in \citet{Zhang23}. Due to resource limitations, we do not include large-scale FMs in this benchmark, e.g., agro-NT and CDSBERT. Besides, we run the evaluation on a subset of GUE, where for each task we randomly select at most 10k samples from the original splits, e.g., training, testing and validation (if any) sets.

\begin{table*}[ht]
    \centering
    \caption{Statistics of tasks in the GUE, these details can be found in Section B.2. from \citet{Zhang23}.}
    \resizebox{.75\linewidth}{!}{
    \begin{tabular}{lccccc}
        \toprule
        \textbf{Task} & \textbf{Metric} & \textbf{Datasets} & \textbf{Training} & \textbf{Validation} & \textbf{Testing} \\ \midrule
        \multirow{3}[1]{*}{Core Promoter Detection} & \multirow{3}[1]{*}{macro F1} & tata & $4,904$ & $613$ & $613$ \\
         &  & notata & $42,452$ & $5,307$ & $5,307$ \\
         &  & all & $47,356$ & $5,920$ & $5,920$ \\ \midrule
        \multirow{3}[1]{*}{Promoter Detection} & \multirow{3}[1]{*}{macro F1} & tata & $4,904$ & $613$ & $613$ \\
         &  & notata & $42,452$ & $5,307$ & $5,307$ \\
         &  & all & $47,356$ & $5,920$ & $5,920$ \\ \midrule
        \multirow{5}[1]{*}{Transcription Factor Prediction (Human)} & \multirow{5}[1]{*}{macro F1} & wgEncodeEH000552 & $32,378$ & $1,000$ & $1,000$ \\
         &  & wgEncodeEH000606 & $30,672$ & $1,000$ & $1,000$ \\
         &  & wgEncodeEH001546 & $19,000$ & $1,000$ & $1,000$ \\
         &  & wgEncodeEH001776 & $27,497$ & $1,000$ & $1,000$ \\
         &  & wgEncodeEH002829 & $19,000$ & $1,000$ & $1,000$ \\ \midrule
        Splice Site Prediction & macro F1 & reconstructed & $36,496$ & $4,562$ & $4,562$ \\ \midrule
        \multirow{5}[1]{*}{Transcription Factor Prediction (Mouse)} & \multirow{5}[1]{*}{macro F1} & Ch12Nrf2\textbackslash iggrab & $6,478$ & $810$ & $810$ \\
         &  & Ch12Zrf384hpa004051\textbackslash iggrab & $5,395$ & $674$ & $674$ \\
         &  & MelJun\textbackslash iggrab & $2,620$ & $328$ & $328$ \\
         &  & MelMafkDm2p5dStd & $1,904$ & $239$ & $239$ \\
         &  & MelNelf\textbackslash iggrab & $15,064$ & $1,883$ & $1,883$ \\ \midrule
        \multirow{10}[1]{*}{Epigenetic Marks Prediction} & \multirow{10}[1]{*}{macro F1} & H3 & $11,971$ & $1,497$ & $1,497$ \\
         &  & H3K14ac & $26,438$ & $3,305$ & $3,305$ \\
         &  & H3K36me3 & $29,704$ & $3,488$ & $3,488$ \\
         &  & H3K4me1 & $25,341$ & $3,168$ & $3,168$ \\
         &  & H3K4me2 & $24,545$ & $3,069$ & $3,069$ \\
         &  & H3K4me3 & $29,439$ & $3,680$ & $3,680$ \\
         &  & H3K79me3 & $23,069$ & $2,884$ & $2,884$ \\
         &  & H3K9ac & $22,224$ & $2,779$ & $2,779$ \\
         &  & H4 & $11,679$ & $1,461$ & $1,461$ \\
         &  & H4ac & $27,275$ & $3,410$ & $3,410$ \\ \midrule
        Covid Variant Classification & macro F1 & Covid & $77,669$ & $7,000$ & $7,000$ \\ \midrule
        \multirow{6}[1]{*}{Enhancer Promoter Interaction} & \multirow{6}[1]{*}{macro F1} & GM12878 & $10,000$ & $2,000$ & $2,000$ \\
         &  & HeLa-S3 & $10,000$ & $2,000$ & $2,000$ \\
         &  & HUVEC & $10,000$ & $2,000$ & $2,000$ \\
         &  & IMR90 & $10,000$ & $2,000$ & $2,000$ \\
         &  & K562 & $10,000$ & $2,000$ & $2,000$ \\
         &  & NHEK & $10,000$ & $2,000$ & $2,000$ \\ \midrule
        \multirow{2}[1]{*}{Species Classification} & \multirow{2}[1]{*}{macro F1} & fungi & $8,000$ & $1,000$ & $1,000$ \\
         &  & virus & $4,000$ & $500$ & $500$ \\ \midrule
    \end{tabular}
    }
    \label{tab:gue_detail}
\end{table*}

The benchmark results on GUE can be found in \pref{tab:gue_results}. While \ourbase\ does not consistently outperform other models across all datasets, it consistently demonstrates top-tier performance despite not being pre-trained on any DNA genome database. These results indicate that, although some FMs are optimised for specific genomic tasks (such as SpliceBERT for splice site detection), \ourbase, which is specifically designed for RNA genomes, shows robust and versatile performance across a variety of tasks. The varying performance across different tasks and species suggests that genomic tasks could benefit from strong generalisability, provided that biological domain knowledge is incorporated into the training of FMs.

\begin{table*}[htbp]
  \centering
  \caption{The performance on GUE for \our\ and baseline FMs, where the results are re-implemented based on our evaluation protocol. The performance for each task is the average macro F1 score in all sub-datasets.}
  \resizebox{.8\linewidth}{!}{
    \begin{tabular}{@{}lccccccc@{}}
    \toprule
          & \multicolumn{7}{c}{\textbf{Model Performance (macro F1 Score)}} \\
    \cmidrule(r){2-8}
    \textbf{Model} & \textbf{Yeast EMP} & \textbf{Mouse TF-M} & \textbf{Virus CVC} & \textbf{Human TF-H} & \textbf{Human PD} & \textbf{Human CPD} & \textbf{Human SSP} \\
    \midrule
    DNABERT-2& $75.85$ & $\mathbf{86.23}$ & $68.90$   & $81.80$  & $90.17$ & $82.57$ & $85.21$ \\
    HyenaDNA & $73.08$ & $73.44$ & $66.37$ & $77.62$ & $91.19$ & $84.31$ & $83.34$ \\
    Caduceus & $73.49$ & $78.18$ & $49.09$ & $79.56$ & $89.13$ & $85.09$ & $81.82$ \\ 
    NT-V2    & $74.93$ & $78.10$  & $59.23$  & $79.12$ & $90.87$ & $84.70$  & $84.13$ \\
    SpliceBERT & $77.66$ & $84.97$ & $56.24$ & $\mathbf{82.77}$ & $\mathbf{92.24}$ & $83.96$ & $\mathbf{93.81}$ \\
    3UTRBERT & $71.89$ & $71.46$ & $68.71$ & $74.85$ & $82.37$ & $\mathbf{90.51}$ & $81.95$ \\
    \ourbase & $\mathbf{78.51}$ & $84.72$ & $\mathbf{74.72}$ & $81.73$ & $90.04$ & $85.22$ & $90.39$ \\
    \bottomrule
    \end{tabular}%
    }
  \label{tab:gue_results}%
\end{table*}%

\section{\strtoseq\ Modelling Case: RNA Design}
\label{app:rna_design}

\subsection{Genetic Algorithm}
\begin{figure}[H]
   \centering
   \includegraphics[width=\linewidth]{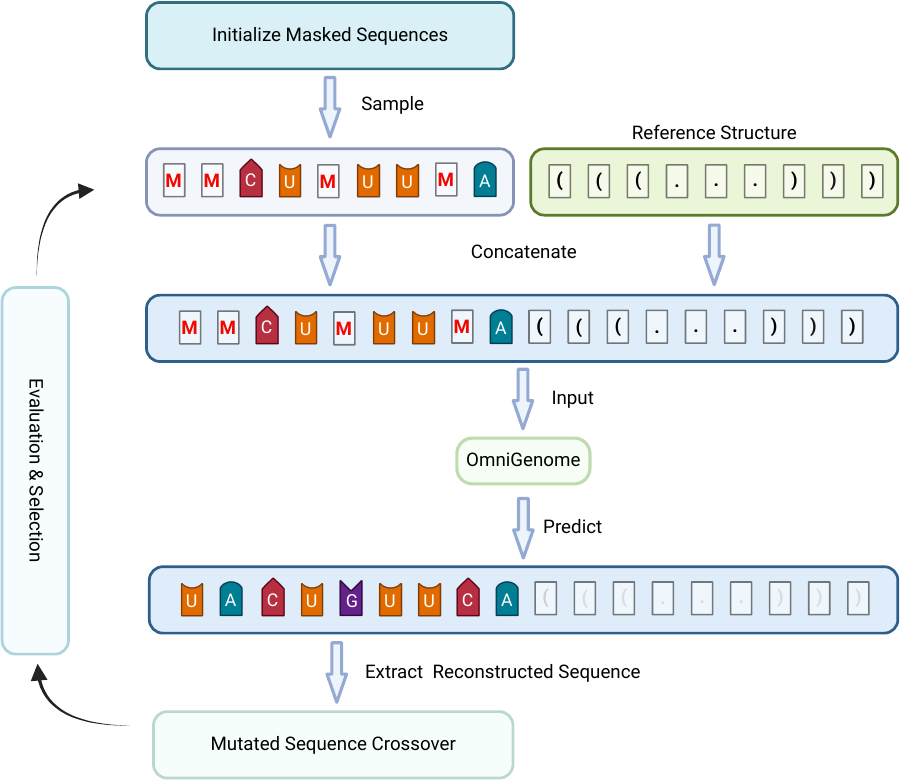}
   \caption{The genetic algorithm used for solving RNA design tasks. `\textcolor{red}{M}' and \textcolor{green}{A} are abbreviations for the mask token and the predicted bases in this mutation operation, respectively. The most effective component in this algorithm is the structure-based sequence reconstruction based on \ourplus.}
   \label{fig:design}
\end{figure}

The working mechanism of our designed genetic algorithm based on \ourplus\ is implemented as the following five-step process:
\begin{itemize}
    \item Step $1$. Given the target RNA secondary structure, we use \our\ to generate a set of candidate sequences $\mathcal{P}=\{\mathbf{s}^i\}_{i=1}^N$.

    \item Step $2$. If the termination criterion is not met, go to Step $3$; otherwise, output the current best sequence $\mathbf{s}^\ast=\text{argmax}_{\mathbf{s}\in\mathcal{P}}f(\mathbf{s})$.

    \item Step $3$. Based on $\mathcal{P}$, use single-point crossover and mutation to generate a population of offspring sequences $\mathcal{O}=\{\tilde{\mathbf{s}}\}_{i=1}^N$.

    \item Step $4$. Combine $\mathcal{P}$ and $\mathcal{O}$ to obtain $\mathcal{S}=\mathcal{P}\bigcup\mathcal{O}$, and use \our\ to predict the corresponding secondary structures of each sequence in $\mathcal{S}$. Evaluate the fitness values of sequences in $\mathcal{S}$.

    \item Step $5$. Sort $\mathcal{S}$ according to the fitness values and preserve the best $N$ sequences to constitute a new $\mathcal{P}$. Return to Step $2$.
\end{itemize}

Note that the fitness value of a sequence $\mathbf{s}$, denoted as $f(\mathbf{s})$, is evaluated as the Hamming distance of the RNA secondary structure predicted by \our\ against the target structure. The above genetic algorithm is not terminated until the sequence for the target RNA secondary structure is identified or the allocated computational budget is exhausted.

\begin{figure*}[t!]
    \centering
    \includegraphics[width=\linewidth]{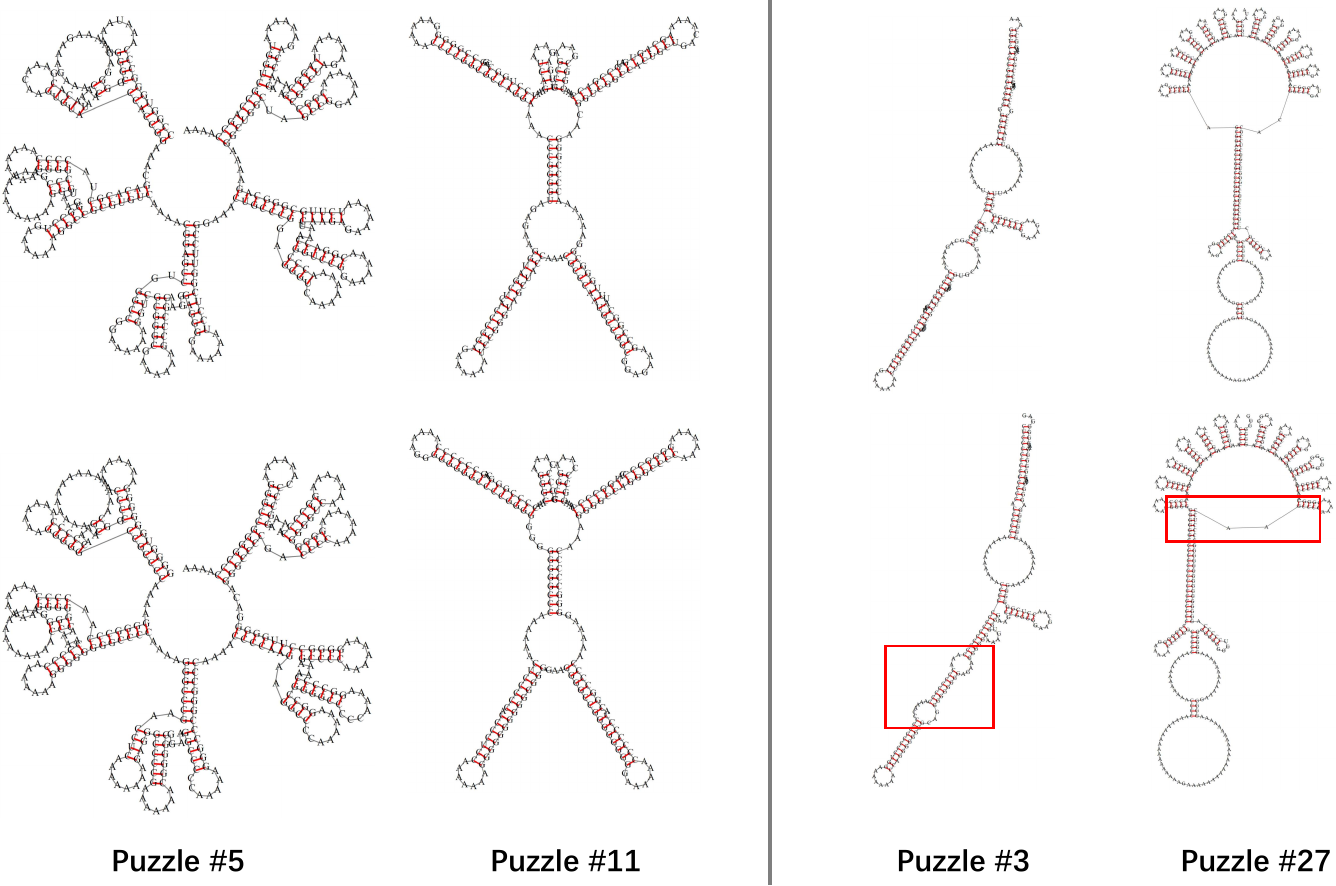}
    \caption{The examples for in-silico RNA design from the EternaV2 design benchmark. Two puzzles (\#5 and \#11) are correctly solved and two puzzles (\#3 and \#27) are incomplete. The top four sequences with structures are the reference solutions, and the bottom sequences are obtained by \ourbaseplus. The structures are derived by ViennaRNA and the \textcolor{red}{red boxes} highlight the difference parts between reference and nearly solved structure.}
    \label{fig:case_design}
\end{figure*}

As demonstrated in the zero-shot experiments in \pref{tab:str2seq}, \ourplus\ models achieve top-tier performance, i.e., \ourplus\ solved $74$ out of $100$ puzzles. We show several complex examples of puzzles from the EternaV2 design benchmark. According to \pref{fig:case_design}, puzzles \#5 and \#11 with approximately $200$+ bases are solved, while these puzzles are challenging to existing FMs. Even for puzzles that are not completely solved, e.g., puzzles \#3 and \#27, \ourbaseplus\ generates very similar structures, where the nucleotide base difference ratio between the designed structure and the target structure is only $\approx 3\%$. This finding indicates the proficiency of \ourplus\ models in solving challenging single-nucleotide resolution genome tasks.

\subsection{\our\ for Long-context RNA Modelling}
\label{app:long-context}
We study RNA genome modelling in this work, and most of the RNA sequences in the wild are short. However, in the case of evaluating the modelling performance on long-context RNA genomes, such as RNA sequences that contain more than $512$ nucleotide bases, we conduct an experiment to verify the SSP performance of \our\ for long sequences. To be more specific, we collected the bpRNA dataset~\cite{DanaeeRWDHH18} from the official website\footnote{\url{https://bprna.cgrb.oregonstate.edu}} and filtered it using the same protocol as the RGB SSP tasks. We re-split this bpRNA dataset based on a length threshold of $512$ bases; i.e., we used the sequences $\geq$ $512$ bases to form the testing set, and the rest of the sequences were split into training and validation sets. This variant of the bpRNA dataset is called bpRNA-L, and the results are available in \pref{tab:bpRNA-L-results}.

\begin{table}[H]
	\centering
	\caption{The performance (F1) on long RNA structure prediction (bpRNA-L) across various approaches.}
	\resizebox{\linewidth}{!}{
	\begin{tabular}{c|cccccc}
		\toprule
		\textbf{Dataset} & ViennaRNA & RNA-BERT & RNA-MSM & RNA-FM & \ourbaseplus \\
		\midrule
		\textbf{bpRNA-L} & $57.73$ & $39.26$ & $39.82$ & $53.44$ & $\mathbf{65.26}$ \\
		\bottomrule
	\end{tabular}%
	}
	\label{tab:bpRNA-L-results}%
\end{table}%

The results indicate that \our\ outperforms existing RNA FMs in long-context RNA structure prediction. Specifically, \ourbaseplus\ achieved the highest F1 score of $65.26$, surpassing other models such as ViennaRNA and RNA-FM. This demonstrates the capability of \our\ in handling complex RNA structures in sequences longer than $512$ bases, reinforcing the importance of sequence-structure alignment in RNA genome modelling.

\section{Pre-training Objective Ablation Experiment}
We have included ablation experiments of pre-training objectives to study their effectiveness. For example, we ablated the \strtoseq\ and \seqtostr\ objectives for the \oursmall\ variant. Due to resource considerations, we only trained each ablation with $1,000$ steps and evaluated the performance on RGB. The results are in \pref{tab:ablation}.

\begin{table}[H]
	\centering
	\caption{Results of the pre-training objective ablation experiments. We only train the following variants for 1k steps due to resource limitations. }
	\resizebox{\linewidth}{!}{
		\begin{tabular}{lcccccc}
			\toprule
			OmniGenome-52M & mRNA  &  SNMD  & SNMR  &  Archive2 &  Stralign &  bpRNA \\
			\cmidrule{2-7}    Ablations & RMSE  & AUC   & F1    & F1    & F1    & F1 \\
			\midrule
			MLM   & $0.7463$ & $53.64$ & $40.95$ & $77.69$ & $93.12$ & $68.99$ \\
			MLM+Str2Seq & $0.7399$ & $56.24$ & $41.83$ & $78.24$ & $93.19$ & $70.11$ \\
			MLM+Seq2str & $0.7421$ & $56.06$ & $41.19$ & $77.73$ & $93.23$ & $69.58$ \\
			MLM+Str2Seq+Seq2str &  $\mathbf{0.7341}$	& $\mathbf{57.20}$ &	$\mathbf{42.81}$ & 	$\mathbf{78.77}$ & 	$\mathbf{93.55}$ &	$\mathbf{71.82}$
			 \\
			\bottomrule
		\end{tabular}%
	}
	\label{tab:ablation}%
\end{table}%
Overall, the results of the ablation experiments show that combining the masked language modelling (MLM) objective with both \strtoseq\ and \seqtostr\ mappings leads to the best performance across all evaluation metrics on the RGB benchmark. The model variant trained with all three objectives (\strtoseq, \seqtostr, and MLM) achieved the lowest RMSE and the highest AUC and F1 scores, indicating that these objectives are complementary and enhance the model's ability to generalise to different RNA genomic tasks.

\section{The OneKP Initiative}
\label{app:1kp}

There has been a variety of FMs utilised in different species, e.g., humans~\citep{Nguyen23,Dalla23}, bacteria~\citep{Nguyen24}, and viruses~\citep{Peng24}, which indicates the effectiveness of pre-trained FMs on multi-species genomics. In this work, we aim to propose an FM for multi-species plant RNA sequence modelling. We leverage the OneKP initiative~\citep{Carpenter19} to address the scarcity of plant RNA data, which contains $1,124$ species of plant transcriptomes. The scale of OneKP enables the development of a more robust and transferable RNA FM.

The $1000$ Plant Transcriptomes Initiative (OneKP) was a comprehensive effort aimed at exploring genetic diversity across the green plant kingdom (Viridiplantae), sequencing the RNA from $1124$ ($1342$ in other versions) samples that represent over $1000$ species, encompassing all major taxa within Viridiplantae. This includes streptophyte and chlorophyte green algae, bryophytes, ferns, angiosperms, and gymnosperms. The initiative's final or capstone publication presents three major analyses: inferring species trees, identifying whole genome duplications, and detecting gene family expansions. These findings are particularly valuable for plant and evolutionary scientists interested in specific gene families, whether their focus is across the entire green plant tree of life or within more narrowly defined lineages.

The sampling strategy for the 1KP was global and collaborative, with samples sourced from a wide range of environments including wild field collections, greenhouses, botanical gardens, laboratory specimens, and algal culture collections. The initiative prioritised the collection of live growing cells, such as young leaves, flowers, or shoots, to ensure a high abundance of expressed genes, though many samples also came from roots and other tissues. RNA extraction was performed using well-established protocols or commercial kits, facilitating the comprehensive analysis of transcribed RNA across this diverse set of species. This monumental effort not only sheds light on plant genetic diversity but also provides a rich data resource for ongoing and future research in plant science and evolutionary biology.

We thank the reviewer’s constructive suggestion. In the camera-ready version, we will refine the broader limitations on the biological and ethical considerations, including the obstacles in in-vivo verifications, and implications in safety screening, etc. Further, we will add a section to discuss the broad potential and prospective applications (e.g., mRNA vaccine design) of our model and elaborate on the out-of-scope concepts and scenarios (e.g., tertiary structure prediction and generalization to other omics data).

\section{Limitations}

Our work has certain limitations, which we are actively addressing in future iterations:

\begin{itemize} 
\item \textbf{Model and Data Scale:} Although our current RNA FM outperforms prior arts in various scenarios, its scale remains small relative to the growing availability of large biological databases and the scaling laws described in \citet{Kaplan20,Hoffmann22,Muennighoff23}. Due to resource constraints, we were unable to pre-train substantially larger models or fully exploit large-scale databases like OneKP. Moving forward, we will focus on training more expansive foundation models to realize improved performance and generalization in both DNA and RNA contexts.

\item \textbf{Sequence Length Constraints:} The modeling length of our current FMs, while adequate for many RNA and DNA tasks (given that RNA sequences are typically shorter than genomes), still may not be sufficient for some applications requiring very long-range modeling. Future directions will include enhancements to handle significantly longer sequences, enabling the model to tackle a wider variety of downstream tasks.

\item \textbf{Biological Verification and Ethical Considerations:} A major challenge in the practical application of our models is the limited availability of cost-effective and ethically sound \textit{in vivo} validation. The utility of predictions generated by our foundation models depends heavily on subsequent experimental studies, which are often expensive, time-consuming, and constrained by ethical and safety standards. Moreover, the use of these models in safety-critical scenarios, such as clinical diagnostics or therapeutic interventions, must be guided by rigorous safety screening, ethical oversight, and adherence to regulatory frameworks.

\item \textbf{Application Scope and Out-of-Scope Extensions:} While we highlight prospective applications such as mRNA vaccine design and RNA-based therapeutics, our current approach does not directly address tasks like tertiary RNA structure prediction or generalization to other omics data types. Integrating these capabilities will require further methodological development, additional data modalities, and possibly more intricate architectures. Similarly, exploring other high-impact scenarios, such as broader gene-editing applications or regulatory RNA element identification, remains future work.

\item \textbf{Implications for Downstream Tasks:} Although we have demonstrated encouraging results in several downstream applications, the complex functional landscapes of RNA biology mean that certain domain-specific tasks may still be challenging. To address these gaps, we plan to incorporate richer biological annotations, explore multi-modal inputs, and collaborate with domain experts to identify critical performance targets and safety checkpoints for biomedical use.
\end{itemize}

\section{Ethics Statement}

This research utilised the open OneKP dataset, which is devoid of human-related privacy concerns. The pre-training sequences used are plant-based genomic data, which could have ecological implications. The following ethical guidelines must be adhered to when using \our:

\begin{itemize}
\item Ensure that the data is used responsibly, with proper attribution and fair compensation to the original sources.
\item Prohibit the use of the model for unethical purposes, such as creating harmful bio-software or designing dangerous RNA structures.
\item The models and findings should contribute to the conservation of plant species and their ecosystems, rather than posing a threat.
\item Adhere to principles of transparency and open science, utilising publicly available datasets and providing comprehensive documentation of methodologies and findings.
\end{itemize}

In conducting this research, we are committed to ethical scientific practices that respect biodiversity and contribute positively to genomic research. We advocate for ongoing dialogue regarding the ethical use of plant RNA sequences and support initiatives that ensure benefits from such research are shared equitably with all stakeholders.

\end{document}